\documentclass{hep99}

\begin{document}

\title{%
\vspace{-3.1cm}
\begin{flushright}
       {\normalsize\tt CERN-TH/99-395}   \\[-0.2cm]
       {\normalsize\tt December 1999}   \\
\end{flushright}
       \vspace{0.7cm}
CP Violation and Rare Kaon Decays$^\dag$}


\author{G Buchalla}
%

\address{Theory Division, CERN, CH-1211 Geneva 23, Switzerland}

\abstract{
We summarize both the study of CP violation with $K$ and $B$ mesons, 
as well as rare decays of kaons, emphasizing recent developments.
The topics discussed include the unitarity triangle, 
$\varepsilon'/\varepsilon$, $K\to\pi\nu\bar\nu$ and other rare
$K$ decays, T-odd asymmetries in kaon physics,
theoretical aspects of CP violation in $B$ decays,
$B\to J/\Psi K_S$, $B\to\pi\pi$, $B\to\pi K$ and inclusive
CP asymmetries.}

\maketitle

\fntext{\dag}{Plenary talk presented at International Europhysics
  Conference on High-Energy Physics (EPS-HEP\,99), Tampere, Finland,
  15--21 July 1999.}

\section{Introduction}

The intensive study of weak interaction processes and the physics
of flavour has proved crucial for our present understanding of
matter at the most fundamental level. 
The idea of strangeness as
an additional quantum number opened extremely fruitful new directions in 
flavour physics. 
At the same time it laid the ground for the quark
model, which in turn provided the basis for the subsequent
development of QCD. The $\theta$--$\tau$ puzzle in kaon decays
suggested the violation of parity, a property now reflected in the
chiral nature of the weak gauge interactions.
The strong suppression of flavour-changing neutral current processes,
as $K_L\to\mu^+\mu^-$ or $K$--$\bar K$ mixing, motivated the
GIM mechanism and the introduction of the charm quark. 
Later, the hint of an
unexpectedly large top-quark mass, infered from $B$--$\bar B$ mixing,
replicated, in a sense, at higher energies, the story of
$K$--$\bar K$ mixing and charm.
Finally, the 1964 discovery of CP violation in $K\to\pi\pi$ decays
already carried the seed of a three-generation Standard Model, ten
years before even the charm quark was found.
These examples illustrate impressively how the careful study of
low energy phenomena may be sensitive to physics at scales much
larger than $m_K$ or $m_B$, and that profound insights can be obtained
by such indirect probes.

Among the various aspects of flavour dynamics the phenomenon of
CP violation is a particular focus of current research.
This fundamental symmetry violation provides us with an absolute
distinction between matter and antimatter, and it is one of the
necessary conditions for a dynamical generation of the baryon asymmetry
in the universe (see the talk by M. Shaposhnikov, these proceedings). 
CP violation is also a very important testing ground
for flavour dynamics in general.

Let us very briefly recall the basic mechanism of CP violation
in the Standard Model. Gauge ($q'_L$) and mass eigenstates ($q_L$)
of up ($q=u$) and down-type ($q=d$) quarks are related by unitary
rotations $U_L$, $D_L$ (of dimension $n$ for $n$ generations)
\begin{equation}\label{uldl}
u'_L=U_L u_L\qquad d'_L=D_L d_L
\end{equation}
Due to a mismatch of $U_L$ and $D_L$, i.e. 
$V=U^\dagger_L D_L\not= 1$, the charged current weak interactions
\begin{equation}\label{lcc}
{\cal L}_{cc}=\frac{g_W}{2\sqrt{2}}V_{ij} 
\,\bar u_i\gamma^\mu(1-\gamma_5)d_j\, W^\dagger_\mu + h.c.
\end{equation}
induce transitions between different families ($i=1,\ldots, n$).
Their coupling strengths are
described by the Cabibbo--Kobayashi--Maskawa (CKM) matrix $V$.
With 1 or 2 generations, $V$ can be chosen to be real. For
3 generations $V$ is parametrized by 3 angles and 1 complex phase,
which is responsible for breaking CP symmetry in eq. (\ref{lcc}).
The CKM matrix has the explicit form
\begin{equation}\label{vckm}
V = \left( \begin{array}{ccc}
           V_{ud} & V_{us} & V_{ub} \\ 
           V_{cd} & V_{cs} & V_{cb} \\ 
           V_{td} & V_{ts} & V_{tb} 
         \end{array} \right)   
\simeq
\end{equation}
\begin{equation}\label{vwo}
\left( \begin{array}{ccc}
           1-\lambda^2/2 & \lambda & A\lambda^3(\varrho- i \eta) \\ 
           -\lambda & 1-\lambda^2/2 & A \lambda^2 \\ 
           A\lambda^3(1-\varrho-i \eta) & -A\lambda^2 & 1
         \end{array} \right) 
\end{equation} 
where the second expression is the useful approximate representation
due to Wolfenstein with the parameters $\lambda$, $A$, $\varrho$
and the complex phase $\eta$.

The replication of fermion generations, quark mixing and CP violation
are striking features of the theory of weak interactions.
While the gauge sector of the theory is well understood and
tested with high precision, the breaking of electroweak symmetry and
its ramifications in flavour physics leave still many questions
unanswered. Detailed and quantitative tests of the flavour sector
have so far remained rather limited. On the other hand the Standard 
Model relates
all phenomena of CP violation to a single complex phase, a constraint
that enables us to put this framework to a decisive test.
Additional CP violating phases occur in many Standard Model extensions 
\cite{GNR,GFG} and could lead to substantial deviations from what we
expect in the Standard Model.

The present talk summarizes recent developments in the field of
CP violation and rare kaon decays. Further aspects of flavour
physics are discussed in the talk by M. Artuso in these proceedings.
After this introduction we give, in section 2, a classification of
CP violation and briefly review the status of the unitarity 
triangle in section 3.
Section 4 is devoted to CP violation and rare decays in the kaon sector,
including $\varepsilon'/\varepsilon$, $K\to\pi\nu\bar\nu$, T-odd
CP asymmetries and a short overview of further rare decays of interest.
In section 5 we discuss CP violation in $B$ decays, summarizing
the theoretical basis and important
applications, $B\to J/\Psi K_S$, $B\to\pi^+\pi^-$, $B\to\pi K$
and the determination of $\gamma$, a selection of other strategies,
and inclusive CP asymmetries. We conclude in section 6.

\section{Classification of CP Violation}

The complex phase in the 3-generation CKM matrix, which is present 
independently of quark phase conventions, violates the CP symmetry
of the Standard Model Lagrangian. As a consequence weak decays 
that would otherwise
be forbidden by CP symmetry may occur, or processes related to each
other by a CP transformation may have different rates. In order for
the CKM phase to manifest itself in such observable asymmetries,
an interference of some sort is in general necessary
to induce a physical phase
{\it difference\/} between the interfering components.
There are various ways in which this general condition may be
realized and it is useful to introduce a classification of
the possible mechanisms. 

In many cases an important role is played by neutral mesons $P$
such as $P=K^0$, $B_d$ or  $B_s$. 
Their characteristic feature is neutral
meson ($P$--$\bar P$) mixing, which is described by a $2\times 2$
Hamiltonian matrix $H=M-i \Gamma/2$.

Following common practice, three classes may then be distinguished:

{\bf a) CP violation in the mixing matrix.}
A typical case is the rate difference
\begin{eqnarray}\label{aksl}
&&\!\!\!\!\!\! 
 \frac{\Gamma(K_L\to\pi^-l^+\nu)-\Gamma(K_L\to\pi^+l^-\bar\nu)}{
         \Gamma(K_L\to\pi^-l^+\nu)+\Gamma(K_L\to\pi^+l^-\bar\nu)}= 
\nonumber \\
&& \frac{|1+\bar\varepsilon|^2-|1-\bar\varepsilon|^2}{
         |1+\bar\varepsilon|^2+|1-\bar\varepsilon|^2}
\simeq 2 Re\bar\varepsilon\simeq
\frac{1}{4}{\rm Im}\frac{\Gamma_{12}}{M_{12}}
\end{eqnarray}
where we have used ($CP\cdot K=-\bar K$)
\begin{equation}\label{klkk}
K_L\sim (1+\bar\varepsilon)K+(1-\bar\varepsilon)\bar K
\end{equation}
Here the lepton charge tags the flavour component ($K$ or $\bar K$)
in the $K_L$. The first equality in (\ref{aksl}) follows immediately.
Note the proportionality of the asymmetry 
to ${\rm Im}(\Gamma_{12}/M_{12})$, which characterizes the effect as
originating in the mixing matrix itself. The sign of the asymmetry,
$2{\rm Re}\bar\varepsilon = 3.3\cdot 10^{-3} > 0$,
allows us to give an absolute definition of positive electric charge.
Similar observables can be constructed for $B$ decays. A peculiarity
of the kaon system is the large hierarchy of lifetimes between the
neutral eigenstates $K_L$ and $K_S$. The long lived $K_L$ state is
singled out by the time evolution itself, which provides a convenient
initial state tag. Explicit initial state flavour tagging
by other means is in general required for neutral $B$ mesons. 

{\bf b) CP violation in the decay amplitude.}
In its simplest form this mechanism can be introduced without
any reference to mixing. In this case the general requirement of
amplitude interference is particularly transparent. Consider the
decay amplitudes
\begin{equation}\label{apf}
A(P\to f)=A_1 e^{i\delta_1}e^{i\phi_1}+A_2 e^{i\delta_2}e^{i\phi_2}
\end{equation}
\begin{equation}\label{apfb}
A(\bar P\to \bar f)=A_1 e^{i\delta_1}e^{-i\phi_1}+
                    A_2 e^{i\delta_2}e^{-i\phi_2}
\end{equation}
where we have assumed two different components $i=1,2$ with
strong phases $\delta_i$ and weak (CKM) phases $\phi_i$. Only the
weak phases change sign in going from $P\to f$ to the CP conjugate
reaction $\bar P\to\bar f$. For the rate difference one finds
\begin{eqnarray}\label{adir}
&&\!\!\!\!\! |A(P\to f)|^2-|A(\bar P\to\bar f)|^2 \nonumber \\ 
&& \sim A_1 A_2 \sin(\delta_1-\delta_2)\sin(\phi_1-\phi_2)
\end{eqnarray}
The conditions for a CP violating asymmetry are, obviously,
the presence of two components in the decay amplitude, as well as both
a strong and a weak phase difference between them. This mechanism is
relevant for neutral mesons, but also for charged mesons
($K^\pm$, $B^\pm$, $\ldots$) or baryons.

{\bf c) CP violation in the interference of mixing and decay.}
In this case the necessary interference arises in an interplay of
decay and mixing. A typical, illustrative example is the
CP violating amplitude
\begin{eqnarray}\label{app0}
&&\!\!\!\!\!\!\!\! A(K_L\to\pi\pi(I=0))\sim \\
&&\!\!\!\!\! (1+\bar\varepsilon)A_0 e^{i\delta_0}-
(1-\bar\varepsilon)A^*_0 e^{i\delta_0}
\sim \bar\varepsilon + i\frac{{\rm Im}A_0}{{\rm Re} A_0}
\equiv\varepsilon \nonumber
\end{eqnarray}
where $\delta_0$ is the strong phase and the weak phase is
included in $A_0$ (as defined after (\ref{epedef})). 
One recognizes the combination of mixing
($\bar\varepsilon$) and decay amplitude ($A_0$) in the observable
$\varepsilon$, the well known CP violation parameter in the kaon
sector. Note that $\varepsilon$ is a physical quantity, in contrast to
the components $\bar\varepsilon$ and $i{\rm Im}A_0/{\rm Re}A_0$,
which are phase convention dependent in the CKM framework.

In analogy to (\ref{app0}) one can consider the decays to 
$\pi^+\pi^-$ and $\pi^0\pi^0$. In the CP symmetry limit $K_L$
would be CP odd and could not decay to the CP even two-pion final
states. Thus the amplitude ratios
\begin{equation}\label{etas}
\eta_{+-}=\frac{A(K_L\to\pi^+\pi^-)}{A(K_S\to\pi^+\pi^-)}\qquad
\eta_{00}=\frac{A(K_L\to\pi^0\pi^0)}{A(K_S\to\pi^0\pi^0)}
\end{equation}
measure CP violation. They can be parametrized as
\begin{equation}\label{eteps}
\eta_{+-}=\varepsilon+\varepsilon'\qquad
\eta_{00}=\varepsilon -2\varepsilon'
\end{equation}
with $\varepsilon$ as in (\ref{app0}) and
\begin{equation}\label{epedef}
\frac{\varepsilon'}{\varepsilon}\simeq
\frac{\omega}{\sqrt{2}|\varepsilon|}
\left(\frac{{\rm Im}A_2}{{\rm Re}A_2}-
      \frac{{\rm Im}A_0}{{\rm Re}A_0}\right)
\end{equation}
where $A(K^0\to\pi\pi(I=0,2))\equiv A_{0,2}e^{i\delta_{0,2}}$
(with strong phases $\delta_{0,2}$ factored out) and
$\omega\equiv{\rm Re}A_2/{\rm Re}A_0\simeq 0.045$.
The smallness of $\omega$ reflects the empirical $\Delta I=1/2$
rule in kaon decays.

If CP violation was only due to mixing, the ratios in (\ref{etas})
were independent of the final state and $\eta_{+-}=\eta_{00}$.
Any difference $\eta_{+-}\not=\eta_{00}$ necessarily
involves CP violation in the decay amplitudes and is measured by
$\varepsilon'$.
Since the ratio $\varepsilon'/\varepsilon$ is real to very good
accuracy, it can be experimentally determined via the
double ratio of rates
\begin{equation}\label{epexdr}
\left|\frac{\eta_{+-}}{\eta_{00}}\right|^2 =
1+6{\rm Re}\frac{\varepsilon'}{\varepsilon}
\end{equation}
In parallel to the classes a) -- c), another terminology is in
use. CP violation in the mixing matrix itself (a)) is refered to
as {\it indirect\/} CP violation. In contrast, CP violation in the
decay amplitude (as in b)) is called {\it direct\/} CP violation.
As we have seen, class c) comprises elements of both the indirect and
the direct effect. In particular, $\varepsilon'/\varepsilon$ is a
measure of direct CP violation.
Completely analogous notions apply to $B$ decays as well, although
some details of the phenomenology are different. We shall come back
to those applications in the $B$ physics sections.

\section{Status of the Unitarity Triangle}

\begin{figure}
\begin{center}
 \vspace{4.0cm}
\includegraphics{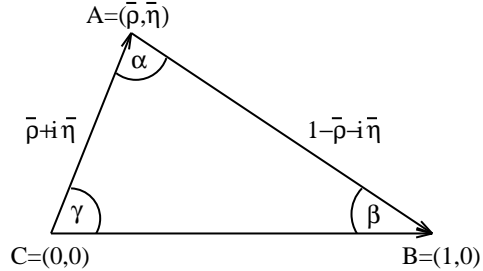}
\end{center} 
\caption{The unitarity triangle in the ($\bar\varrho$, $\bar\eta$) plane.
    \label{fig:utabc} }
\end{figure}

\begin{figure}
\begin{center}
 \vspace{3.5cm}
\includegraphics{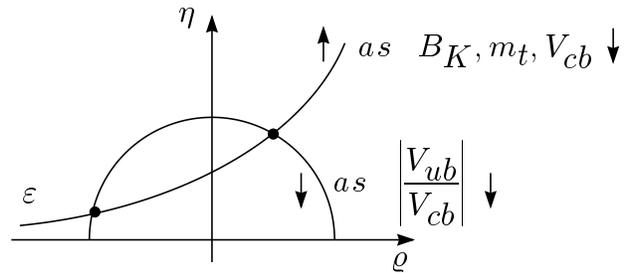}
\end{center} 
\caption{The $\varepsilon$-constraint on the unitarity triangle.
    \label{fig:uteub} }
\end{figure}

\begin{figure}
\begin{center}
 \vspace{5.5cm}
\includegraphics{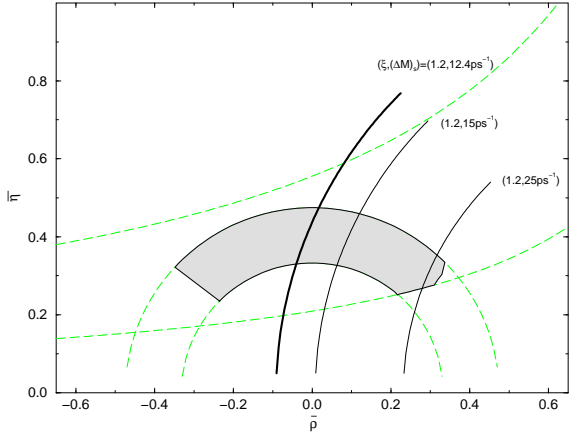}
\end{center} 
\caption{The allowed region (shaded) in the ($\bar\varrho$, $\bar\eta$) 
plane, combining information from $\varepsilon$, $|V_{ub}/V_{cb}|$
and including the constraint from $\Delta M_d$. The independent 
constraint from the lower limit on $\Delta M_s/\Delta M_d$
excludes the region to the left of the curves labeled with 
$\Delta M_s$ in the plot. $\xi\simeq 1.2$ measures $SU(3)$ breaking
in the hadronic matrix elements of 
$B_d-\bar B_d$ versus $B_s-\bar B_s$ mixing. The plot is taken from
\cite{AJB99} where further particulars can be found.
Similar analyses have been presented in \cite{PRS,ALO,MEL,CPS}.
    \label{fig:utnum} }
\end{figure}

The CKM unitarity relation
\begin{equation}\label{utrel}
V^*_{ub}V_{ud}+V^*_{cb}V_{cd}+V^*_{tb}V_{td}=0
\end{equation}
defines a triangle in the complex plane. Rescaling the sides
by $|V^*_{cb}V_{cd}|$, and keeping subleading terms in the
Wolfenstein expansion in $\lambda$ \cite{BLO}, the unitarity triangle is
displayed in the plane of $\bar\varrho=\varrho(1-\lambda^2/2)$
and $\bar\eta=\eta(1-\lambda^2/2)$ (fig. \ref{fig:utabc}).
This representation is useful since it emphasizes the least known
CKM quantities $\bar\varrho$ and $\bar\eta$.
Fig. \ref{fig:utabc} also defines the CKM angles
$\alpha$, $\beta$ and $\gamma$.
At present the most important constraints on ($\bar\varrho$, $\bar\eta$),
the apex of the unitarity triangle, are coming from kaon CP violation
($\varepsilon$), semileptonic $B$ decays ($|V_{ub}/V_{cb}|$) and
$B-\bar B$ mixing ($\Delta M_d$ and $\Delta M_d/\Delta M_s$,
constraining $|V_{td}|$ and $|V_{td}/V_{ts}|$, respectively).
In principle, $\varepsilon$ and $|V_{ub}/V_{cb}|$ are sufficient to
determine the unitarity triangle, as illustrated in fig. \ref{fig:uteub}.
The characteristic dependence on the relevant input parameters
$m_t$, $V_{cb}$, $|V_{ub}/V_{cb}|$ and the hadronic matrix element
$B_K$ is indicated in this plot. Note that the consistency of the
Standard Model, i.e. intersection of the two curves, 
requires all four parameters 
to be sufficiently {\it large}. In particular the large value
of the top-quark mass $\bar m_t(m_t)=167\,{\rm GeV}$ is important
for the -- nontrivial -- consistency of the Standard Model description.
The theory of $\varepsilon$ and $B-\bar B$ mixing at 
next-to-leading order in QCD
is described in \cite{HN,BJW}.
The current status of the unitarity triangle obtained from a
detailed numerical analysis is shown in fig. \ref{fig:utnum}
(see also M. Artuso, these proceedings).

\section{Kaons: CP Violation and Rare Decays}

\subsection{$\varepsilon'/\varepsilon$}

Until very recently all CP violation effects ever observed in the
laboratory were different manifestations of a single quantity,
the parameter $\varepsilon$ describing indirect CP violation in the
neutral kaon system. This $\varepsilon$-effect was first seen 1964
in $K_L\to\pi^+\pi^-$ and subsequently in a few further decay channels
of $K_L$, namely $\pi^0\pi^0$, $\pi l\nu$, $\pi^+\pi^-\gamma$ and
finally $\pi^+\pi^-e^+e^-$. As we have seen, the experimental
value of $\varepsilon$ is well compatible with the Standard Model. 
Nevertheless
it is clear that this quantity alone gives insufficient information
to provide us with a decisive test.
Complementary and qualitatively new insight can be gained by
measuring direct CP violation in $K\to\pi\pi$, i.e. 
$\varepsilon'/\varepsilon$. 
In superweak scenarios \cite{GNR,WOL} $\varepsilon'$ is predicted
to vanish, while it is generically nonzero in the Standard Model.
Considerable efforts were therefore invested over the years
to determine $\varepsilon'/\varepsilon$ in experiment, but the
situation as to whether $\varepsilon'/\varepsilon$ was zero or not
had remained inconclusive.
Two measurements, about ten years old, characterize the 
experimental status before 1999
\begin{equation}\label{epex}
{\rm Re}\frac{\varepsilon'}{\varepsilon}=
\left\{ \begin{array}{ll}
        (23\pm 6.5)\cdot 10^{-4} & \mbox{CERN NA31} \\
        (7.4\pm 5.9)\cdot 10^{-4} & \mbox{FNAL E731}
        \end{array}
\right.
\end{equation}
Earlier this year the successor experiments at both laboratories
have announced their first new results
\cite{KTEVEPE,NA48EPE} (see also H. Nguyen (KTeV) and
S. Palestini (NA48), these proceedings) 
\begin{equation}\label{epex99}
{\rm Re}\frac{\varepsilon'}{\varepsilon}=
\left\{ \begin{array}{ll}
        (18.5\pm 7.3)\cdot 10^{-4} & \mbox{CERN NA48} \\
        (28.0\pm 4.1)\cdot 10^{-4} & \mbox{FNAL KTeV}
        \end{array}
\right.
\end{equation}
An average of all four numbers gives
\begin{equation}\label{epexav}
{\rm Re}\frac{\varepsilon'}{\varepsilon}=(21.2\pm 4.6)\cdot 10^{-4}
\end{equation}
These new experimental achievments (\ref{epex99}) 
are among the major highlights of this year in particle physics.
They establish direct CP violation for the first time, confirming
the earlier evidence from NA31.
Both NA48 and KTeV are still ongoing and a third experiment,
KLOE, at the Frascati $\Phi$-factory has just started.
$\varepsilon'/\varepsilon$ will therefore be known with
still higher precision ($\sim 10^{-4}$) in the future.

A detailed theoretical interpretation of these results is
a very difficult task. Although important progress has been
made over the years, the calculation of $\varepsilon'/\varepsilon$
still suffers from considerable uncertainties related to the low energy
dynamics of QCD. 
Before presenting the theoretical status of $\varepsilon'/\varepsilon$,
we begin by recalling the essential features of the calculational
framework.
The expression (\ref{epedef}) for $\varepsilon'/\varepsilon$ may
also be written as
\begin{equation}\label{epea02}
\frac{\varepsilon'}{\varepsilon}=-\frac{\omega}{\sqrt{2}
  |\varepsilon|\mbox{Re} A_0}\left(\mbox{Im} A_0-
  \frac{1}{\omega}\mbox{Im} A_2\right)
\end{equation}
$\mbox{Im} A_{0,2}$ are calculated from the general low energy
effective Hamiltonian for $\Delta S=1$ transitions \cite{BBL}. 
Including
electroweak penguins this Hamiltonian involves ten different
operators and one has
\begin{equation}\label{ima02}
\mbox{Im} A_{0,2}=-\mbox{Im}\lambda_t \frac{G_F}{\sqrt{2}}
 \sum^{10}_{i=3} y_i(\mu)\langle Q_i\rangle_{0,2}
\end{equation}
Here $y_i$ are Wilson coefficients, $\lambda_t=V^*_{ts}V_{td}$
and
\begin{equation}\label{kppq}
\langle\pi\pi(I=0,2)|Q_i|K^0\rangle\equiv
\langle Q_i\rangle_{0,2}e^{i\delta_{0,2}}
\end{equation}
For the purpose of illustration we keep only the numerically
dominant contributions and write
\begin{equation}\label{epeapr}
\frac{\varepsilon'}{\varepsilon}=
\frac{\omega G_F}{2|\varepsilon|\mbox{Re} A_0}\mbox{Im}\lambda_t
\left(y_6\langle Q_6\rangle_0-\frac{1}{\omega}y_8\langle Q_8\rangle_2
+\ldots\right)
\end{equation}
$Q_6$ originates from gluonic penguin diagrams and $Q_8$ from
electroweak contributions. The matrix elements of $Q_6$ and
$Q_8$ can be parametrized by bag parameters $B_6$ and $B_8$ as
\begin{equation}\label{q6me}
\langle Q_6\rangle_0 =-\sqrt{24}
\left[\frac{m_K}{m_s(\mu)+m_d(\mu)}\right]^2 m^2_K(f_K-f_\pi)\cdot B_6
\end{equation}
\begin{equation}\label{q8me}
\langle Q_8\rangle_2\simeq\sqrt{3}
\left[\frac{m_K}{m_s(\mu)+m_d(\mu)}\right]^2 m^2_K f_\pi\cdot B_8
\end{equation}
that is
\begin{equation}\label{q68b}
\langle Q_6\rangle_0\sim \left(\frac{m_K}{m_s}\right)^2 B_6
\qquad
\langle Q_8\rangle_2\sim \left(\frac{m_K}{m_s}\right)^2 B_8
\end{equation}
$B_6=B_8=1$ corresponds to the factorization assumption for the
matrix elements, which holds in the large $N_C$ limit of QCD.
\\
$y_6\langle Q_6\rangle_0$ and $y_8\langle Q_8\rangle_2$
are positive numbers. The value for $\varepsilon'/\varepsilon$
in (\ref{epeapr}) is thus characterized by a cancellation of
competing contributions. Since the second contribution is an
electroweak effect, suppressed by $\sim\alpha/\alpha_s$ compared
to the leading gluonic penguin $\sim\langle Q_6\rangle_0$,
it could appear at first sight that it should be altogether
negligible for $\varepsilon'/\varepsilon$. However, a number of
circumstances actually conspire to systematically enhance the
electroweak effect so as to render it a sizable contribution:
\begin{itemize}
\item
Unlike $Q_6$, which is a pure $\Delta I=1/2$ operator,
$Q_8$ can give rise to the $\pi\pi(I=2)$ final state and thus
yield a non-vanishing $\mbox{Im} A_2$.
\item
The ${\cal O}(\alpha/\alpha_s)$ suppression is largely compensated
by the factor $1/\omega\approx 22$ in (\ref{epeapr}), reflecting the
$\Delta I=1/2$ rule.
\item
$-y_8\langle Q_8\rangle_2$ gives a negative contribution to
$\varepsilon'/\varepsilon$ that strongly grows with $m_t$
\cite{FR,BBH}. For the realistic top mass value it can be substantial.
\end{itemize}

In the following we will summarize current theoretical analyses
of $\varepsilon'/\varepsilon$, point out the major sources
of uncertainty and briefly describe some recent developments.

To display the most important ingredients for theoretical predictions,
it is useful to consider an approximate numerical formula for
$\varepsilon'/\varepsilon$ \cite{AJBK99}
\begin{eqnarray}\label{epeapp}
\frac{\varepsilon'}{\varepsilon} &\approx&
13 \mbox{Im}\lambda_t\left(
  \frac{\Lambda^{(4)}_{\overline{MS}}}{340\,{\rm MeV}}\right)
 \left[\frac{110\,{\rm MeV}}{m_s(2\,{\rm GeV})}\right]^2 
\nonumber \\
&&\!\!\!\!\!\!\!\!\!\!\!\!\!\!\!\!\! 
   \cdot\left[B_6(1-\Omega_{IB})-0.4 B_8
    \left(\frac{m_t(m_t)}{165\,{\rm GeV}}\right)^{2.5}\right]
\end{eqnarray}
This expression is a convenient approximation for the purpose of
illustration, but it should not be used for a detailed quantitative
analysis.

The Wilson coefficients $y_i$ in (\ref{ima02}), (\ref{epeapr})
have been calculated at next-to-leading order
\cite{BJLW,CFMR}. The short-distance part in (\ref{epeapp})
is therefore quite well under control. 
The largest theoretical uncertainties come from the hadronic matrix
elements, in particular from $\langle Q_6\rangle_0$ and
$\langle Q_8\rangle_2$, which are parametrized by $B_6$, 
$B_8$ and $m_s$. Another hadronic quantity entering the analysis is
$\Omega_{IB}$, representing the effect of isospin breaking
in the quark masses ($m_u\not= m_d$).
An estimate of $\Omega_{IB}=0.25\pm 0.08$ is reported in
\cite{AJBK99} and has been used in most analyses of
$\varepsilon'/\varepsilon$. 

These quantities parametrize nonperturbative effects of QCD that
are difficult to compute in practice. They are not fundamental
parameters of the Standard Model, but could be expressed, in principle,
in terms of the latter, that is $\Lambda_{QCD}$, $m_u$, $m_d$, $m_s$.
(Note that $m_s$ entering (\ref{epeapp}) does not represent the
complete physical $m_s$-dependence of $\varepsilon'/\varepsilon$,
as the kaon mass $m^2_K=r m_s$ is always kept fixed; $m_s$ is
introduced indirectly as a conventional way to rewrite the value of
the quark condensate $\sim r$. The condensate appears in the
factorized matrix elements of $Q_{6,8}$, which are products of
matrix elements of two (pseudo-) scalar currents.)
In principle the dependence of $\langle Q_6\rangle_0$ and
$\langle Q_8\rangle_2$ on the strange-quark mass $m_s$ need
not be made explicit if the matrix elements are calculated directly
within a nonperturbative approach such as lattice QCD.
However, $m_s$ enters naturally in the large-$N_c$ limit of QCD,
where the matrix elements factorize and $B_6=B_8=1$ are valid
exactly. Since the Wilson coefficients $y_{6,8}$ are already of
${\cal O}(1/N_c)$, the leading-order approximation of $B_{6,8}$ 
in large-$N_c$ is formally consistent with a next-to-leading order 
treatment of the
full decay amplitude, known to be important for the CP conserving
real parts \cite{BBG}.
Also, the dependence of $\langle Q_6\rangle_0$ and
$\langle Q_8\rangle_2$ on the renormalization scale $\mu$ is
carried almost entirely by the quark masses in (\ref{q6me}), 
(\ref{q8me}). $B_6$ and $B_8$ are thus practically $\mu$-independent.
This feature is convenient for comparing various analyses, where the
$B$-factors are obtained at different scales.
For these reasons the parametrization of $\langle Q_6\rangle_0$ and
$\langle Q_8\rangle_2$ in terms of $B_6$, $B_8$ and $m_s$
is still a useful convention.

The value of the strange-quark mass has received considerable
attention recently. Representative
numbers are $m_s(2\,{\rm GeV})=(110\pm 20){\rm MeV}$ from
quenched lattice simulations \cite{GKL} and
$m_s(2\,{\rm GeV})=(124\pm 22){\rm MeV}$ from QCD sum rules
\cite{QCDS}, the latter result quoted here as the average reported 
in \cite{BBGJJLS}.
Recent quenched lattice calculations obtain for $m_s(2\,{\rm GeV})$
the values
$(121\pm 13){\rm MeV}$ \cite{BEC}, 
$(106\pm 7){\rm MeV}$ \cite{JLQCD},
$(97\pm 4){\rm MeV}$ \cite{UKQCD} and
$(105\pm 5){\rm MeV}$ \cite{QCDSF}. Unquenching is expected to
lower these values by about $15{\rm MeV}$ (\cite{SRY} and refs. therein).
The tendency towards relatively small $m_s$ had already been
emphasized in \cite{LANL,FNAL}.
For a review on lattice determinations of the strange-quark mass
see e.g. \cite{SRY} and the contribution of H. Wittig in
these proceedings.
On the other hand it has been argued on the basis of dispersion
relations that $m_s$ obeys a lower bound, typically
$m_s(2\,{\rm GeV}) \stackrel{>}{_\sim} 100\,{\rm MeV}$
\cite{LRT}. The framework employed is similar to the one in QCD sum rule
calculations, but less assumptions are used to avoid model dependence.
An alternative method determines $m_s$ from 
Cabibbo-suppressed hadronic $\tau$ decays, where the latest analysis finds
$m_s(2\,{\rm GeV}) =(114\pm 23)\,{\rm MeV}$ \cite{PPR}
(see also \cite{ALMS,CKP} for earlier results).

Small values of the strange-quark mass, as the ones indicated 
particularly by lattice calculations, contribute to increasing
the theoretical values for $\varepsilon'/\varepsilon$, at fixed
$B_6$, $B_8$.

Various groups have presented analyses of $\varepsilon'/\varepsilon$.
Their findings are summarized in table 1.
\begin{table*}\label{tab:epe}
\begin{center}
\caption{Summary of recent theoretical results for
$\varepsilon'/\varepsilon$, together with the hadronic quantities
$B_6$ and $B_8$ used in the analysis.
$B_6$ and $B_8$ are taken to be in the NDR scheme; they are
approximately scale independent. The values for $B_6$ and $B_8$
of the Rome group are rescaled to correspond to a nominal
strange-quark mass of $m_s(2\,{\rm GeV}) =110\,{\rm MeV}$.
(MC) denotes a Gaussian treatment of uncertainties for the
experimental input, and (S) the more conservative scanning of
all parameters within their assumed ranges
(see the quoted references for more details).}
\begin{tabular}{llll} 
\br
$B_6$ & $B_8$ & $\varepsilon'/\varepsilon[10^{-4}]$ & reference \\
\mr
$1.0\pm 0.3$ & $0.8\pm 0.2$ & $7.7^{+6.0}_{-3.5}$ (MC) 
    & Munich \cite{BBGJJLS} \\
$1.0\pm 0.3$ & $0.8\pm 0.2$ & $1.1\to 28.8$ (S) 
    & Munich \cite{BBGJJLS} \\
$0.7\pm 0.7$ & $0.7\pm 0.1$ & $6.7^{+9.2}_{-8.5}\pm 0.4$ (MC)
    & Rome \cite{CFGLM} \\
$0.7\pm 0.7$ & $0.7\pm 0.1$ & $-10\to 30$ (S)
    & Rome \cite{CFGLM} \\
$0.91\pm 0.19$ & $B_6/1.72$ & $2.1\to 26.4$ (S) 
    & Dortmund \cite{HKPS} \\
$1.33\pm 0.25$ & $0.77\pm 0.02$ & $7\to 31$ (S) 
    & Trieste \cite{BFE} \\
$1.0$ & $1.0$ & $-3.2\to 3.3$ (S) & Dubna \cite{BBLM} \\
\br
\end{tabular}
\end{center}
\end{table*}

All results are based on the effective Hamiltonian with Wilson
coefficients calculated in \cite{BJLW,CFMR}. There are slight
variations in the choice of input parameters, as those entering
the determination of ${\rm Im}\lambda_t$ from $\varepsilon$.
The most important differences are in the treatment of hadronic
matrix elements.
All groups use values of $m_s$ close to the representative
range \cite{BBGJJLS}
\begin{equation}\label{msrep}
m_s(2\,{\rm GeV}) =(110\pm 20)\,{\rm MeV}
\end{equation}

Several approaches have been used to obtain estimates for
$B_6$ and $B_8$. Unfortunately none of them can be considered
fully satisfactory at present and in particular the error
bars are hard to quantify with confidence.

The leading order large-$N_c$ limit gives $B_6=B_8=1$
\cite{BBG}.
More recent estimates based on the large-$N_c$ approach and
using a simultaneous chiral ($p^2$) and $1/N_c$ expansion
find \cite{HKPSB}
\begin{equation}\label{b6b8d}
B_6=0.72 \mbox{--} 1.10\qquad
B_8=0.42 \mbox{--} 0.64
\end{equation}
These estimates include the corrections to the matrix elements
of ${\cal O}(p^2)$ and ${\cal O}(p^0/N_c)$ (the leading 
term ${\cal O}(p^0)$ is nonvanishing only in the case of $B_8$).
A particular term at higher order (${\cal O}(p^2/N_c)$)
is identified that would enhance $B_6$ to about $1.5$
\cite{HKPS,SHS}. In view of other contributions and systematic
uncertainties of the approach such a conclusion has to be taken
with caution and appears premature at present.

The Trieste group \cite{BFE} employs a chiral quark model to evaluate
hadronic quantities. They find
\begin{equation}\label{b6b8t}
B_6=1.07 \mbox{--} 1.58\qquad
B_8=0.75 \mbox{--} 0.79
\end{equation}
The model is able to fit the $\Delta I=1/2$ rule and leads
to enhanced values for $B_6$, however, its relation to QCD is
not entirely clear.

Ultimately, the most reliable results should come from a
first-principles lattice calculation.
For the matrix elements under discussion the lattice method is,
however, at present still affected by large systematic uncertainties
(for instance, only $\langle\pi|Q_i| K\rangle$ is simulated directly 
and lowest order chiral perturbation theory is used to obtain
the required matrix element with two pions in the final state;
also quenching introduces uncertainties that are hard to quantify).
In particular no reliable results seem to be available
for $B_6$ \cite{CFGLM}, which is plagued by large corrections
from lattice perturbation theory \cite{PK}. A nonperturbative
lattice-continuum matching procedure is therefore required \cite{MPSTV}.
For $B_8$ one finds from lattice QCD computations
\cite{KGS,CDGMTV,GBS}
\begin{equation}
B_8=0.69\mbox{--} 1.06
\end{equation}

As is evident from table 1, the uncertainties in
theoretical calculations of $\varepsilon'/\varepsilon$ are
very large. Still, there is a tendency for results with
bag parameters $B_{6,8}$ in the vicinity of $1$ to yield
estimates of $\varepsilon'/\varepsilon$ below the experimental
values.

This situation has encouraged additional theoretical efforts
aimed at a better understanding of various aspects of the
low-energy hadronic dynamics entering $\varepsilon'/\varepsilon$.
Let us briefly mention some recent activities.

The authors of \cite{BPR} also use a large-$N_c$ framework,
supplemented by the extended Nambu--Jona-Lasinio (ENJL) model
to interpolate low-energy QCD, and quote $B_6=2.2\pm 0.5$,
which would comfortably accommodate the sizable experimental numbers.
Unfortunately also this result is not entirely free of model-dependent
input.

An interesting approach is discussed in \cite{DGO},
where $B_8$ is obtained from a dispersive analysis in the
strict chiral limit. It is found that
$B_8=1.11\pm 0.16\pm 0.23$ in the NDR scheme and assuming
$m_s(2\,{\rm GeV})+m_d(2\,{\rm GeV})=100\,{\rm MeV}$
(see also J. Donoghue, these proceedings).
Similar ideas are presented in \cite{KPDR} in the context of 
the large-$N_c$ limit.

A promising new method in lattice QCD employs domain-wall fermions
to ensure good chiral properties on the lattice (see
A. Soni, these proceedings).
The Riken-BNL-Columbia collaboration \cite{RBC} has performed
an exploratory study finding very large and negative $B_6$,
which translates into large and negative 
$\varepsilon'/\varepsilon=
(-3.3\pm 0.3(\mbox{stat})\pm 1.6(\mbox{syst}))\cdot 10^{-2}\eta$.
Taken literally, the central value would imply a striking
disagreement with experimental results. However, the error
bars are still very large and the numbers obtained have still to be 
considered preliminary, awaiting further scrutiny.

The effect of isospin breaking due to the $m_u$--$m_d$ mass
difference, parametrized by $\Omega_{IB}$, has been reconsidered
in \cite{GV}. In this paper it is argued that new sources of
isospin breaking at order ${\cal O}(p^4)$  in chiral perturbation 
theory might have a sizable impact, possibly leading even to negative
$\Omega_{IB}$ and a resulting enhancement of $\varepsilon'/\varepsilon$.
The issue was subsequently also addressed in \cite{EMNP}, 
where the contribution of
$\pi^0-\eta$ mixing to $\Omega_{IB}$ is obtained to be 
$\Omega^{\pi^0\eta}_{IB}=0.16\pm 0.03$ within
chiral perturbation theory to ${\cal O}(p^4)$.  

Another mechanism that would lead to larger values of $B_6$ than
in conventional estimates is discussed in \cite{KNS}. In this
paper it is proposed that the effect of the $\sigma$-resonance,
specific to the $I=0$ channel of the two-pion final state,
could increase $B_6$ and $\varepsilon'/\varepsilon$ considerably.
This effect corresponds to a resummation of a class of
two-pion rescattering diagrams.

A complete, exact calculation of the matrix elements
$\langle\pi\pi(I)|Q_i|K^0\rangle$ would automatically yield
the correct final state interaction (FSI) phase factors
$\exp(i\delta_I)$ multiplying $\langle Q_i\rangle_I$ in
(\ref{kppq}).
Current methods of estimating the matrix elements, however, 
do not correctly reproduce these phases, which are known experimentally.
The phases obtained are either zero or much smaller than the
empirical values. A general discussion of FSI in the context of
$\varepsilon'/\varepsilon$ was very recently given in \cite{PP},
starting from the Omn\`es solution of a dispersion relation for
the decay amplitudes. It was argued that FSI enhance the 
$I=0$ channel, and hence $B_6$, and suppress $I=2$ contributions.
This would bring the central value of $\varepsilon'/\varepsilon$
closer to the data. The framework discussed in \cite{PP} is
certainly interesting. It will be important to see the
same features emerge also within the context of a complete
and explicit calculation of the matrix elements.

There is a general consensus that it is impossible, at present,
to conclude a significant discrepancy with the Standard Model
in the measurement of $\varepsilon'/\varepsilon$, taking into
account the substantial hadronic uncertainties.
Nevertheless, contributions from non-standard physics could
indeed affect the value of $\varepsilon'/\varepsilon$, and the
very sizable experimental figures for this quantity have reinforced
the interest in exploring New-Physics scenarios.
A brief summary of these investigations, most of which
focus on supersymmetric models, and further references
can be found in \cite{AJBK99,UN} (see also Y.-L. Wu, these
proceedings).
Typically the most likely sources of New Physics in
$\varepsilon'/\varepsilon$ are $Z$-penguin and chromomagnetic
penguin contributions. They correspond to operators of
dimension 4 and 5, respectively, and could be quite naturally,
by dimensional counting, larger than higher dimensional contributions
from New Physics \cite{BCIRS} (see \cite{KN} for an alternative
scenario in supersymmetry).

Additional general discussions of the theory of $\varepsilon'/\varepsilon$
may be found, for instance, 
in the recent review articles \cite{AJBK99,UN,MJ}.

The clear and unambiguous experimental determination of
$\varepsilon'/\varepsilon$ has demonstrated the existence of
direct CP violation. This result thus confirms the qualitative
expectation of nonzero $\varepsilon'$ characteristic of
the CKM mechanism and rules out superweak scenarios where 
$\varepsilon'/\varepsilon=0$. 
The new measurements have also reinforced theoretical efforts to
gain an improved understanding of the complicated hadronic dynamics.
Further progress may be anticipated, although a good accuracy
of the computations is likely to remain a serious theoretical
challenge in the future.
The tantalizing situation of an intriguing experimental result
whose interpretation is hampered by large theoretical uncertainties,
lends further motivation to look for other observables with better
theoretical control. Examples are provided by certain
rare decays of kaons, which will form the subject of the
following section.

\subsection{$K\to\pi\nu\bar\nu$}\label{sec:kpnn}

\begin{figure}
 \vspace{3cm}
\includegraphics{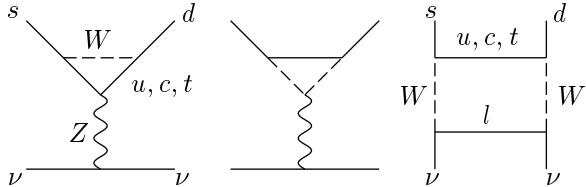}
 \caption{\it
      Leading order electroweak diagrams contributing to
      $K\to\pi\nu\bar\nu$ in the Standard Model.
    \label{fig:kpnn} }
\end{figure}

In this paragraph we focus on the rare decays $K^+\to\pi^+\nu\bar\nu$
and $K_L\to\pi^0\nu\bar\nu$, which are particularly promising.
In these modes the loop-induced FCNC transition $s\to d$ is probed by
a neutrino current, which couples only to heavy gauge bosons
($W$, $Z$), as shown in fig. \ref{fig:kpnn}. 
Correspondingly, the GIM pattern of the
$\bar s\to\bar d\nu\bar\nu$ amplitude has, roughly speaking, the
form 
\begin{equation}\label{asdnn}
A(\bar s\to\bar d\nu\bar\nu)\sim\lambda_i m^2_i
\end{equation} 
summed over $i=u$, $c$, $t$ 
($\lambda_i=V^*_{is}V_{id}$). The power-like
mass dependence strongly enhances the short-distance contributions, 
coming from the heavy flavours $c$ and $t$.
(This is to be contrasted with the logarithmic mass dependence
of the photonic penguin, important for $\bar s\to\bar d e^+e^-$.)
The short-distance dominance has, then, two crucial
consequences. First,
the transition proceeds through an effectively local
$(\bar sd)_{V-A}(\bar\nu\nu)_{V-A}$
interaction.
Second, because that local interaction is semileptonic,
the only hadronic matrix element required,
$\langle\pi|(\bar sd)_V|K\rangle$, can be obtained from 
$K^+\to\pi^0l^+\nu$ decay using isospin.
As a result $K\to\pi\nu\bar\nu$ is calculable
completely and with exceptional theoretical control.
While $K^+\to\pi^+\nu\bar\nu$ receives both top and charm
contributions, $K_L\to\pi^0\nu\bar\nu$ probes direct CP violation
\cite{LI} and is dominated entirely by the top sector.

The $K\to\pi\nu\bar\nu$ modes have been studied in great detail
over the years to quantify the degree of theoretical precision.
Important effects come from short-distance QCD corrections.
These were computed at leading order in \cite{DDG}.
The complete next-to-leading order  calculations \cite{BB123,MU,BB99}
reduce the theoretical uncertainty in these decays to
$\sim 5\%$ for $K^+\to\pi^+\nu\bar\nu$ and $\sim 1\%$ for
$K_L\to\pi^0\nu\bar\nu$.
This picture is essentially unchanged when further effects
are considered, including isospin breaking in the relation of
$K\to\pi\nu\bar\nu$ to $K^+\to\pi^0l^+\nu$ \cite{MP},
long-distance contributions
\cite{RS,HLLW}, the CP-conserving effect in $K_L\to\pi^0\nu\bar\nu$
in the Standard Model \cite{RS,BI} and two-loop electroweak 
corrections for large $m_t$ \cite{BB7}.
The current Standard Model predictions for the branching ratios are
\cite{AJB99}
\begin{eqnarray}\label{bkpnn}
B(K^+\to\pi^+\nu\bar\nu) &=& (0.8\pm 0.3)\cdot 10^{-10} \\
B(K_L\to\pi^0\nu\bar\nu) &=& (2.8\pm 1.1)\cdot 10^{-11}
\end{eqnarray}

The study of $K\to\pi\nu\bar\nu$ can give crucial information
for testing the CKM picture of flavor mixing. This information is
complementary to the results expected from $B$ physics and is much
needed to provide the overdetermination of the unitarity triangle
necessary for a real test. 
Let us briefly illustrate  some specific opportunities.

$K_L\to\pi^0\nu\bar\nu$ is probably the best probe of the Jarlskog
parameter $J_{CP}\sim {\rm Im}\lambda_t$, the invariant measure
of CP violation in the Standard Model \cite{BB6}. 
For example a $10\%$ measurement
$B(K_L\to\pi^0\nu\bar\nu)=(3.0\pm 0.3)\cdot 10^{-11}$ would directly
give ${\rm Im}\lambda_t=(1.37\pm 0.07)\cdot 10^{-4}$, a remarkably 
precise result.

Combining $10\%$ measurements of both $K_L\to\pi^0\nu\bar\nu$
and $K^+\to\pi^+\nu\bar\nu$ determines the unitarity
triangle parameter $\sin 2\beta$ with an uncertainty of about
$\pm 0.07$, comparable to the precision obtainable for
the same quantity from CP violation in $B\to J/\Psi K_S$
before the LHC era. 

A measurement of $B(K^+\to\pi^+\nu\bar\nu)$ to $10\%$ accuracy
can be expected to determine $|V_{td}|$ with similar precision.

As a final example, using only information from the ratio
of $B_d-\bar B_d$ to $B_s-\bar B_s$ mixing,
$\Delta M_d/\Delta M_s$, one can derive a stringent and
clean upper bound \cite{BB99}
\begin{eqnarray}\label{kpnxs}
&&\!\!\!\!\!\!\!\!\!\! B(K^+\to\pi^+\nu\bar\nu) \\
&&\!\!\!\!\!\!   < 0.4\cdot 10^{-10}
\left[P_{charm}+A^2 X(m_t)\frac{r_{sd}}{\lambda}
\sqrt{\frac{\Delta M_d}{\Delta M_s}}\right]^2 \nonumber
\end{eqnarray}
Note that the $\varepsilon$-constraint or $V_{ub}$ with their
theoretical uncertainties are not needed here. Using
$V_{cb}\equiv A\lambda^2<0.043$, $r_{sd}<1.4$ (describing
SU(3) breaking in the ratio of $B_d$ to $B_s$ mixing matrix elements)
and $\sqrt{\Delta M_d/\Delta M_s}<0.2$, gives the bound
$B(K^+\to\pi^+\nu\bar\nu)< 1.67\cdot 10^{-10}$, which can be confronted
with future measurements of $K^+\to\pi^+\nu\bar\nu$ decay.
Here we have assumed
\begin{equation}\label{delms}
\Delta M_s > 12.4\,{\rm ps}^{-1}
\end{equation}
corresponding to the world average presented at this conference.
A future increase in this lower bound will strengthen the bound
in (\ref{kpnxs}) accordingly.
Any violation of (\ref{kpnxs}) will be a clear signal of physics
beyond the Standard Model.

Indeed, the decays $K\to\pi\nu\bar\nu$, being highly suppressed
in the Standard Model, could potentially be very sensitive to
New-Physics effects.
This topic has been addressed repeatedly in the recent literature
\cite{GN,NW,HHW,BRS,BSIK,CI,BCIRS}.
Most discussions have focussed in particular on general
supersymmetric scenarios \cite{NW,BRS,BSIK,CI,BCIRS}.
Large effects are most likely to occur via enhanced
$Z$-penguin contributions. This is expected because the
$\bar sdZ$ vertex is a dimension-4 operator 
(allowed by the breaking of electroweak symmetry) in the
low-energy effective theory, where the heavy degrees of freedom
associated with the New Physics have been integrated out.
The corresponding $Z$-penguin amplitude for
$\bar s\to\bar d\nu\bar\nu$ will thus be $\sim 1/M^2_Z$, much
larger than the New Physics contribution of dimension 6
scaling as $\sim 1/M^2_S$, if we assume that the scale of
New Physics $M_S\gg M_Z$.
It has been pointed out in \cite{CI} that, in a generic
supersymmetric model with minimal particle content and R-parity
conservation, the necessary
flavour violation in the induced $\bar sdZ$ coupling is
potentially dominated by double LR mass insertions related
to squark mixing. This mechanism could lead to sizable enhancements
still allowed by known constraints.
An updated discussion is given in \cite{BCIRS}.
Typically, enhancements over the Standard Model branching
ratios could be up to a factor of 10 (3) for
$K_L\to\pi^0\nu\bar\nu$ ($K^+\to\pi^+\nu\bar\nu$) within this framework.

\begin{figure}
 \vspace{6cm}
\includegraphics{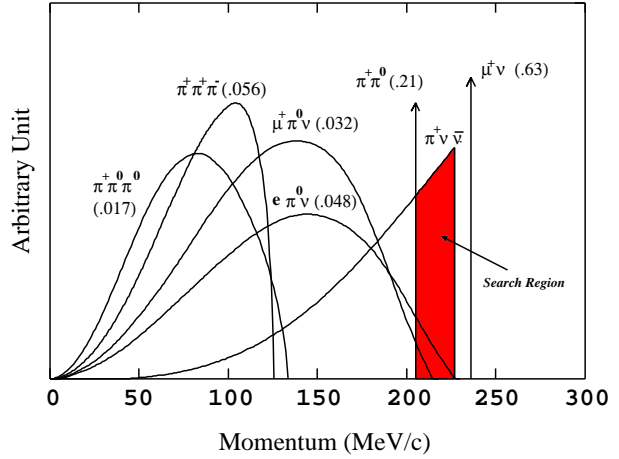}
 \caption{\it
      Momentum spectrum of charged particles from stopped
      $K^+$ decays (from \cite{KOM}).
    \label{fig:kpnspec} }
\end{figure}

In the experimental quest for $K\to\pi\nu\bar\nu$ an
important step has been accomplished by Brookhaven experiment
E787, which observed a single, but very clean candidate event
for $K^+\to\pi^+\nu\bar\nu$ in 1997. This event is practically
background free and corresponded to a branching fraction
of
$B(K^+\to\pi^+\nu\bar\nu)=(4.2^{+9.7}_{-3.5})\cdot 10^{-10}$
\cite{ADL}.
The experiment is very challenging and requires extremely good
control over possible background (see fig. \ref{fig:kpnspec}).
Charged pions from $K^+$ decays at rest are analyzed
in momentum. The signal pions have a continuous momentum
spectrum, which distinguishes them from the monochromatic
spectra of the major background modes
$K^+\to\pi^+\pi^0$ (where the $\pi^0$ escapes detection)
and $K^+\to\mu^+\nu$ (where $\mu^+$ is misidentified
as $\pi^+$). Excellent photon rejection capability and
distinction
of $\pi^+$ versus $\mu^+$, by requiring
observation of the full $\pi^+\to\mu^+\to e^+$ decay sequence,
are essential elements of the experiment.
E787 is continuing and has very recently released an
updated result, based on about $2.5$ times the data underlying
the previous measurement. In addition to the single, earlier
event, no new signal candidates are observed, which translates
into \cite{SAT}
\begin{equation}\label{kpnn99}
B(K^+\to\pi^+\nu\bar\nu)=
\left(1.5^{+3.5}_{-1.3}\right)\cdot 10^{-10}\qquad \mbox{BNL E787}
\end{equation}
The experiment is still ongoing and will be followed by a
successor experiment, E949 \cite{E949}, at Brookhaven. 
Recently, a new experiment, CKM \cite{CKM}, has been proposed to measure 
$K^+\to\pi^+\nu\bar\nu$ at the Fermilab Main Injector,
studying $K$ decays in flight.
Plans to investigate this process also exist at KEK for
the Japan Hadron Facility (JHF) \cite{JHFS}.

The neutral mode, $K_L\to\pi^0\nu\bar\nu$, is currently
pursued by KTeV. The present upper limit reads \cite{KLPNTEV}
\begin{equation}\label{klpn99}
B(K_L\to\pi^0\nu\bar\nu) < 5.9\cdot 10^{-7}\qquad\mbox{KTeV}
\end{equation}
For $K_L\to\pi^0\nu\bar\nu$ a model
independent upper bound can be infered from the experimental result
on $K^+\to\pi^+\nu\bar\nu$ \cite{GN}.
It is given by $B(K_L\to\pi^0\nu\bar\nu)< 4.4 B(K^+\to\pi^+\nu\bar\nu)
< 2\cdot 10^{-9}$. At least this sensitivity will have to be achieved
before New Physics is constrained with $B(K_L\to\pi^0\nu\bar\nu)$.
Concerning the future of $K_L\to\pi^0\nu\bar\nu$ experiments, 
a proposal exists at Brookhaven (BNL E926) to measure this decay at 
the AGS with a sensitivity of ${\cal O}(10^{-12})$ \cite{E926}.
There are furthermore plans to pursue this mode with comparable
sensitivity at Fermilab \cite{KAMI} and KEK \cite{JHFI}.
The prospects for $K_L\to\pi^0\nu\bar\nu$ at a $\phi$-factory are
discussed in \cite{BCI}.

It will be very exciting to follow the development and outcome of
the ambitious projects aimed at measuring $K^+\to\pi^+\nu\bar\nu$
and  $K_L\to\pi^0\nu\bar\nu$. Their goal is highly worthwhile.

\subsection{T-odd Asymmetries}

The time reversal operation T is closely connected with
the CP transformation. If CPT is conserved, which by the
CPT theorem is true for quantum field theories under very general
assumptions, then the known violation of CP symmetry implies
T violation. It is thus of great interest to study T-odd 
observables and to measure those effects in the laboratory.
With the first observation of such T-odd asymmetries,
reported recently by CPLEAR at CERN and KTeV at Fermilab,
the subject has received renewed attention.

CPLEAR studies kaons from proton-antiproton annihilation at
low energies. This experiment measured the following T-odd
rate asymmetry (Kabir test) \cite{CPLEAR}
\begin{eqnarray}\label{cplear}
&&\!\!\!\!\!\!\!\!\!\frac{R[\bar K(0)\to K(t)]-R[K(0)\to \bar K(t)]}{
R[\bar K(0)\to K(t)]+R[K(0)\to \bar K(t)]}\nonumber \\
&&\ \ \ \ \ \ \ \ =(6.6\pm 1.6)\cdot 10^{-3}
\end{eqnarray}
observing, e.g., $p\bar p\to K^+\pi^-\bar K^0$ and the subsequent
decay via mixing $\bar K^0\to K^0\to\pi^- e^+\nu$. The charged
kaon provides the initial tag (time $0$), the charged lepton the
final tag (time $t$) of the flavour of the neutral kaon.
From conventional CP violation and assuming CPT, the asymmetry
is expected to be
$4{\rm Re}\bar\varepsilon=(6.6\pm 0.1)\cdot 10^{-3}$,
in excellent agreement with (\ref{cplear}).

The KTeV collaboration investigates kaon decays in a
high-energy fixed-target experiment. They studied the rare decay
$K_L\to\pi^+\pi^- e^+e^-$.
The amplitude for this decay mode receives two contributions with 
different CP properties. First, a $K_L\to\pi^+\pi^-$ decay, 
where the charged pion legs radiate a virtual photon that
produces the $e^+e^-$ pair.
This mechanism, refered to as the bremsstrahlung contribution,
is clearly CP violating because of the underlying $K_L\to\pi\pi$
transition. The second component is a $M1$ direct emission amplitude,
where the virtual photon and the pions originate from the same vertex.
This contribution is CP conserving.
There are other contributions, but these are the most important ones.
The interference of the two amplitudes induces an asymmetry in
the angular distribution of the decay rate, $d\Gamma/d\phi$,
where $\phi$ is the angle between the ($\pi^+\pi^-$)- and
the ($e^+ e^-$) decay plane. The distribution can be written as
\begin{equation}\label{dgdp}
\frac{d\Gamma}{d\phi}=\Gamma_1\cos^2\phi +\Gamma_2\sin^2\phi
+\Gamma_3\sin\phi\cos\phi
\end{equation}
The last term $\sim\Gamma_3$ describes the asymmetry.
Using 
$\vec z=(\vec p_{\pi^+}+\vec p_{\pi^-})/|\vec p_{\pi^+}+\vec p_{\pi^-}|$,
the unit vector in the direction of the total pion momentum,
and the normals to the pion (electron) decay planes, 
$\vec n_\pi$ ($\vec n_e$), one has
\begin{equation}\label{sphi}
\sin\phi=(\vec n_\pi\times \vec n_e)\cdot\vec z
\end{equation}
Thus, $\sin\phi$ is both CP-odd and T-odd. The interference term
in (\ref{dgdp}) can be isolated by forming the asymmetry
\begin{equation}\label{appee}
|{\cal A}|=\left| \frac{1}{\Gamma}\int d\phi\frac{d\Gamma}{d\phi}
{\rm sign}(\sin\phi\,\cos\phi)\right|
\end{equation}
The existence of this asymmetry was pointed out in
\cite{SWH} and predicted to be $|{\cal A}|\approx 14\%$.
The subject was subsequently also discussed in \cite{EWS}.
It is interesting to note that a CP asymmetry of this size can
be generated in $K_L$ decay, where typical effects are at the
permille level. The reason is a relative dynamical suppression of the
CP conserving component that allows it to interfere more effectively
with the small CP violating amplitude.

KTeV has measured the asymmetry and the branching ratio and found
\cite{KTEVTO}
\begin{equation}\label{aktev}
|{\cal A}|=13.6\pm 2.5 \pm 1.2 \%
\end{equation}
\begin{equation}\label{bktev}
B(K_L\to\pi^+\pi^- e^+e^-)=(3.32\pm 0.14\pm 0.28)\cdot 10^{-7}
\end{equation}
Both asymmetry and branching ratio are in full agreement
with the predictions \cite{SWH}.
Preliminary results were also presented by NA48 at CERN. They 
likewise agree with expectations (J. Nassalski, these proceedings):
\begin{equation}\label{ana48}
|{\cal A}|=20 \pm 5 \% \ \ ({\rm MC} 22\%)\ \   [\mbox{with cuts}]
\end{equation}
\begin{equation}\label{bna48}
B(K_L\to\pi^+\pi^- e^+e^-)=(2.90\pm 0.15)\cdot 10^{-7}
\end{equation}

Under the assumption of CPT invariance, both the Kabir test
and the angular asymmetry in $K_L\to\pi^+\pi^- e^+e^-$ are
equivalent to the well-known indirect CP violation effect in the 
neutral kaon system. (The possibility to extract information
on direct CP violation from  $K_L\to\pi^+\pi^- e^+e^-$ has been
discussed in \cite{SWH,EWS}. Unfortunately the expected signals are
very small.)
The significance of both experiments is that they demonstrate
explicitly the existence of T-odd asymmetries in nature,
anticipated from CP violation on the basis of the CPT theorem.

A somewhat different question is the problem of testing T violation
directly, without assuming CPT symmetry. In general, the existence
of a T-odd correlation in a decay process does not by itself imply
T violation, which requires the interchange of initial and final
states in constructing the asymmetry. This condition is, for instance,
not fulfilled in $K_L\to\pi^+\pi^-e^+e^-$ decay. On the other hand,
it holds apparently for the Kabir test. A potential problem for
a direct test of T violation in this latter case is the need to use
tagging via the decay $K\to\pi e\nu$. Independent measurments can
however be used to avoid this loophole \cite{CPLEAR}. These issues are
further discussed in \cite{WOT,EM,AKLP,BS,SVL}.

There are other T-odd observables that have been proposed in the
literature. A typical example is the transverse muon polarization
\begin{equation}\label{ptmu}
P^T_\mu=\langle\hat s_\mu\cdot
\frac{(\vec p_\mu\times \vec p_\pi)}{|\vec p_\mu\times \vec p_\pi|}
\rangle
\end{equation}
in $K^+\to\pi^0\mu^+\nu$ decay. A nonvanishing polarization
could arise from the interference of the leading, standard
W-exchange amplitude with a charged-higgs exchange contribution
involving CP violating couplings. $P^T_\mu$ is therefore an interesting
probe of New Physics \cite{GK} with conceivable effects
of $P^T_\mu\sim 10^{-3}$. Planned experiments could reach a sensitivity
of $P^T_\mu\sim 10^{-4}$ \cite{ABE,DIW}.
Independently of CP or T violation a nonvanishing $P^T_\mu$ can in
principle be induced by final state interactions (FSI).
(Note that $P^T_\mu\not= 0$ is not forbidden by CP or T symmetry,
although it can be induced when these symmetries are violated.)
In the case of $K^+\to\pi^0\mu^+\nu$ FSI phases arise only at
two loops in QED and are very small ($\sim 10^{-6}$) \cite{ZHI}; 
they would be much larger in $K^0\to\pi^-\mu^+\nu$.
We remark that, in contrast, the angular asymmetry in
$K_L\to\pi^+\pi^-e^+e^-$ discussed above cannot be generated by
FSI alone since it is forbidden by CP invariance. 

Another T-odd observable of interest is the transverse muon
polarization in $K^+\to\mu^+\nu\gamma$ (see \cite{HI}
for a recent discussion). 
Related topics are also addressed in the contribution
of C.-Q. Geng, these proceedings.

\subsection{Further Results on Rare $K$ Decays}

The field of rare $K$ decays is very rich in opportunities.
They range from studies of chiral perturbation theory as a framework
to describe the low-energy dynamics of QCD in weak decays
($K_S\to\gamma\gamma$, $K_L\to\pi^0\gamma\gamma$, 
$K_S\to\pi^0e^+e^-$, $K_L\to l^+l^-\gamma$, $\ldots$), over
tests of flavour dynamics at the weak scale
($K\to\pi\nu\bar\nu$, $K_L\to\pi^0e^+e^-$, $K_L\to\mu^+\mu^-$,
$\ldots$), to searches for exotic phenomena, such as lepton
flavour violation, with potential sensitivity to scales of
several $100\,{\rm TeV}$ ($K_L\to\mu e$, $K\to\pi\mu e$).
A number of useful review articles exists in the literature
on the subject of rare $K$ decays and kaon CP violation
\cite{LV,WW,RW,BR,DI}.

Without going into details we shall summarize here some recent
experimental results that are relevant to various important aspects
of this large and fruitful field of research.
(More information can be found in the talks of parallel session 4
at this conference.)

The updated results from KTeV include the branching ratio limits
\begin{eqnarray}\label{ktevkpl}
B(K_L\to\pi^0e^+e^-) &<& 5.6\cdot 10^{-10}\quad
  (\sim 5\cdot 10^{-12}) \nonumber \\
B(K_L\to\pi^0\mu^+\mu^-) &<& 3.4\cdot 10^{-10}\quad
  (\sim 1\cdot 10^{-12}) \nonumber 
\end{eqnarray}
where the Standard Model expectation is shown in brackets.

NA48 presented measurements of the radiative decays
\begin{eqnarray}\label{na48rad}
B(K_L\to e^+e^-\gamma) &=& (1.05\pm 0.02\pm 0.04)\cdot 10^{-5} 
\nonumber \\
B(K_L\to e^+e^-\gamma\gamma) &=& (5.82\pm 0.27\pm 0.49)\cdot 10^{-7} 
\nonumber 
\end{eqnarray}

The current limits on kaon decays violating lepton family number are
\begin{eqnarray}\label{klfv}
B(K_L\to\mu e) &<& 4.7\cdot 10^{-12}\quad
  \mbox{BNL E871} \nonumber \\
B(K^+\to\pi^+\mu^+ e^-) &<& 4.8\cdot 10^{-11}\quad
  \mbox{BNL E865} \nonumber \\
B(K_L\to\pi^0\mu e) &<& 3.2\cdot 10^{-9}\quad
  \mbox{KTeV} \nonumber 
\end{eqnarray}
from experiments at Brookhaven and Fermilab.
As a byproduct of the search for $K_L\to\mu e$,
BNL E871 has observed the mode $K_L\to e^+e^-$, finding \cite{E871}
\begin{equation}\label{klee}
B(K_L\to e^+e^-)=(8.7^{+5.7}_{-4.1})\cdot 10^{-12}
\end{equation}
This result testifies to the extraordinary precision achievable
in kaon experiments and is noteworthy as the smallest branching
ratio ever observed.
Theoretically this decay is determined by long distance contributions.
Because those are dominated by calculable large logarithms
$\ln(m_K/m_e)$, theoretical estimates \cite{VAL,GDP}
are nevertheless relatively reliable
and gave predictions in agreement with (\ref{klee}).

\section{CP Violation in $B$ Decays}

\subsection{Theoretical Basis}

$B$ meson decays offer a large array of exciting possibilities
to expand our knowledge of flavour physics and to complement
what we can learn from kaon studies.
The large number of available channels makes this field rich
and complex, but provides us at the same time with multiple
opportunities to combine different pieces of information and to
extract the underlying physics.
Of special interest is the question of CP violation in the
$B$ sector, which is most promisingly addressed using nonleptonic
modes.

The theoretical basis for a discussion of hadronic $B$ decays is
given by the low-energy effective Hamiltonian 
(low-energy with respect to $M_W$). 
To be specific, let us consider strangeness-changing 
($\Delta S=1$)
$b$ decays, which describe for instance $B\to\pi K$, but also
$B\to J/\Psi K_S$. In this case the effective Hamiltonian has the form
\begin{eqnarray}\label{heffb}
&&{\cal H}_{eff} = \\
&&\frac{G_F}{\sqrt{2}}V^*_{ub}V_{us}
\left( C_1 Q^u_1+C_2 Q^u_2+\sum_p C_p Q_p\right)\nonumber \\
&&+
\frac{G_F}{\sqrt{2}}V^*_{cb}V_{cs}
\left( C_1 Q^c_1+C_2 Q^c_2+\sum_p C_p Q_p\right)\nonumber
\end{eqnarray}
Here the $C_i$ are Wilson coefficients, containing the short-distance
physics from scales $\mu > \mu_b={\cal O}(m_b)$, while the $Q_i$
are local four-quark operators, whose matrix elements comprise
the contributions from scales $\mu < \mu_b$. The operators
have the flavour form $Q^u_i\sim(\bar bu)(\bar us)$,
$Q^c_i\sim(\bar bc)(\bar cs)$ and for the penguin operators
$Q_p\sim(\bar bs)(\bar qq)$ ($q=u$, $d$, $s$, $c$, $b$),
and come in different Dirac and colour structures.
The Hamiltonian for $\Delta S=0$ transitions (describing
for instance $B\to\pi\pi$) is similar to (\ref{heffb}),
with an obvious change of flavour labels in the operators and
CKM quantities ($s\leftrightarrow d$).

\begin{figure}
 \vspace{3cm}
\includegraphics{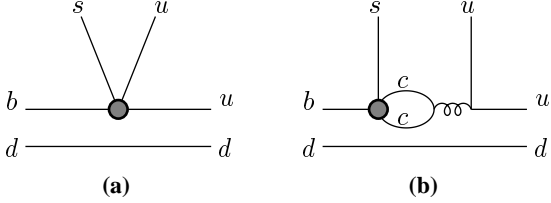}
 \caption{\it
   Typical matrix elements of local four-quark operators for
   $B\to\pi^- K^+$: Leading order tree-level matrix element (a),
   penguin contraction involving a charm-loop (b).
    \label{fig:bme} }
\end{figure}

Typical diagrams for the matrix elements are illustrated in
fig. \ref{fig:bme}. 
In writing (\ref{heffb}) the CKM combination $V^*_{tb}V_{ts}$
has been eliminated using CKM unitarity, so that the GIM cancellation
is already manifest. The amplitudes to be derived from (\ref{heffb})
are clearly seen to exhibit ingredients necessary for CP violation.
There are two components with, in general, different weak and strong
phases, for instance
\begin{equation}\label{gatap}
A(B\to\pi^- K^+)=e^{i\gamma} A_T+A_P
\end{equation}
where the term proportional to $A_T$ ($A_P$) corresponds to the
second (third) line in (\ref{heffb}), and we have used
standard phase conventions with ${\rm arg}(V^*_{ub}V_{us})=\gamma$, 
${\rm arg}(V^*_{cb}V_{cs})\approx 0$.
Here we have made the weak phases explicit, while strong phases are
still contained in $A_{T,P}$.
The components $\exp(i\gamma)A_T$ and $A_P$ are often refered to as the
tree- and the penguin amplitude, respectively.

Eq. (\ref{heffb}) allows us already to read off some gross features
for the amplitudes of a given decay. In general $C_p\ll C_{1,2}$.
Also $V^*_{ub}V_{us}\sim\lambda^4\ll V^*_{cb}V_{cs}\sim\lambda^2$.
Now if, for example, the final state contains $u$-quarks, 
as $B\to\pi^- K^+$,
$Q^u_{1,2}$ contribute at leading order, $Q^c_{1,2}$ only through
loops. Then the second line of (\ref{heffb}) can compete with the
third line, despite the CKM suppression of the former. For
$B\to J/\Psi K_S$, on the contrary, $Q^c_{1,2}$ give the largest
contribution and the term with $V^*_{ub}V_{us}$ is entirely negligible.

The Wilson coefficients are well under control theoretically. They
are known at next-to-leading order in renormalization group improved
QCD \cite{BJLW,CFMR} (for a review see \cite{BBL}).
The main theoretical problem are the matrix elements,
e.g. $\langle\pi K|Q_i|B\rangle$ or $\langle\pi\pi|Q_i|B\rangle$, 
which involve complicated nonperturbative QCD dynamics.
In order to solve (or to circumvent) this problem, the following
possibilities could be distinguished.

First, the hadronic matrix elements may simply cancel in the
observable of interest, as is the case for the CP asymmetry in
$B\to J/\Psi K_S$.
Second, the uncertainties can sometimes be eliminated or
reduced by combining
various channels, exploiting SU(2) or SU(3) flavour symmetry.
A general framework for parametrizing the hadronic amplitudes in a 
consistent way in terms of the possible Wick contractions has
been developed in \cite{CFMS,BSIB}. This could be a useful
starting point for the systematic incorporation of other
approaches or approximations.
Finally, one can try to compute the QCD dynamics of the decay
process directly, at least in a certain limit.

We will come back to applications of the first two possibilities
below and turn now to theoretical approaches towards computing
explicitly the matrix elements of two-body nonleptonic $B$ decays.

A simple approach, which has been widely used in phenomenological
studies \cite{BSW,NS}, is the {\it naive factorization\/} of
matrix elements, schematically
\begin{eqnarray}\label{nfac}
&& \!\!\!\!\!\!\!\langle\pi^+\pi^-|(\bar ub)_{V-A}(\bar du)_{V-A}
|\bar B\rangle \to
\\
&& \langle\pi^+|(\bar ub)_{V-A}|\bar B\rangle\cdot
 \langle\pi^-|(\bar du)_{V-A}|0\rangle\nonumber
\end{eqnarray}
for the example of $B\to\pi\pi$. The justification for this
procedure has been less clear (see also
\cite{AGCT} for applications and \cite{BSNF} for a critical
discussion). 
An obvious issue is the proper
scheme and scale dependence of the matrix elements of four-quark
operators, which is needed to cancel the corresponding dependence
in the Wilson coefficients. Clearly, this dependence is lost
in naive factorization as the factorized currents are scheme
independent objects. In many cases the factorization can be justified
in the large-$N_c$ limit of QCD \cite{BGR}, 
but this approximation is often
too crude for a reliable phenomenology. In any case one would prefer
not to rely exclusively on this argument.
A different qualitative justification for factorization has been
given by Bjorken \cite{BJ}. It is based on the {\it colour
transparency\/} of the hadronic environment for the highly energetic
pion emitted in the decay of a $B$ meson (the $\pi^-$ in the
above example, which is being created from the vacuum).
Formally this is related to the decoupling of soft gluons from
the small-size colour-singlet object that the emitted pion represents.
This aspect of cancellation of soft IR divergences was also
addressed in \cite{DG} in the context of $B\to D\pi$ decay
employing a {\it large-energy effective theory (LEET)\/} framework.
The important issues of collinear singularities and the formation
of the boundstate pion were however not treated in that paper.
The idea to compute decays such as $B\to\pi\pi$ in perturbative QCD
has been pursued using a {\it hard-scattering approach\/} in 
\cite{SHB}. The result of this study raised however several
questions that remained open. For instance, the $B\to\pi$ transition
form factor was assumed to be dominated by hard gluon exchange,
leading to unrealistically small rates. The role of nonfactorizable
contributions was not explicitly addressed.
A hard-scattering framework was also applied by other groups
\cite{SW,DJK,BME,YL}, but without providing a clear conceptual basis.
The question of separating model-dependent from model-independent
ingredients in the analysis, or issues as the scheme dependence of 
matrix elements were left unanswered.
A further method to gain dynamical insight into exclusive hadronic
matrix elements uses {\it QCD sum rule techniques\/} to estimate
nonfactorizable effects. It has been first discussed for
$B\to D\pi$ decays \cite{BSF} (see also \cite{IHA} for class II decays)
and was later applied to $B\to J/\Psi K$ \cite{KR}. Although
the validity of the sum rule implies certain restrictions on the
applicability of this approach, it can be expected to give useful
information in various cases.
Ideas to address nonfactorizable effects in $B\to D\pi$ by means
of an operator product expansion are discussed in \cite{DP}.

A new, systematic approach, going beyond previous 
attempts, was recently formulated in \cite{BBNS}.
There it was proposed that factorization can be justified within
QCD to leading order in the heavy-quark limit for a large class
of two-body hadronic $B$ decays.
The statement of {\it QCD factorization\/} in the case of
$B\to\pi\pi$, for instance, can be schematically written
\begin{eqnarray}\label{qfac}
&&\!\!\!\!\!\! A(B\to\pi\pi)= \\
&&\langle\pi|j_1|B\rangle\, \langle\pi|j_2|0\rangle\cdot
\left[1+{\cal O}(\alpha_s)+
{\cal O}\left(\frac{\Lambda_{QCD}}{m_B}\right)\right]\nonumber
\end{eqnarray}
Up to corrections suppressed by $\Lambda_{QCD}/m_B$ the amplitude is
calculable in terms of simpler hadronic objects: It factorizes,
to lowest order in $\alpha_s$, into matrix elements of bilinear
quark currents ($j_{1,2}$) that are expressed by form factors
such as $f_\pi$ and $f_+(0)$ \cite{KR,BA}. 
To higher order in $\alpha_s$, but still to leading order in
$\Lambda_{QCD}/m_B$, there are nonfactorizable corrections, which are
however governed by hard gluon exchange. They are therefore again
calculable in terms of few universal hadronic quantities,
such as the pion wave function $\Phi_\pi(x)$.
This treatment of hadronic $B$ decays is based on the analysis of
Feynman diagrams in the heavy-quark limit, utilizing consistent
power counting to identify the leading contributions. The framework
is very similar in spirit to more conventional applications  of
perturbative QCD in exclusive hadronic processes with a large 
momentum transfer, as the pion electromagnetic form factor 
\cite{FJ,ER,LB}.
The assumptions used, namely the demonstration of a separation of
hard and soft physics to leading power in a high energy process
on the basis of Feynman diagrams, is very general and underlies
basically all practical applications of perturbative QCD.
In this sense the approach of \cite{BBNS} constitutes a general,
systematic and QCD-based framework to analyze nonleptonic $B$ decays.
The new approach may be viewed as a consistent formalization of 
Bjorken's colour transparency argument \cite{BJ}. In addition
the method includes, for $B\to\pi\pi$, the hard
non-factorizable spectator interactions, penguin contributions
and rescattering effects. As a corollary, one finds that strong
rescattering phases are either of ${\cal O}(\alpha_s)$, and
calculable, or power suppressed. In any case they vanish therefore
in the heavy-quark limit.
QCD factorization is valid for cases where the emitted particle
(the meson created from the vacuum in the weak process, as opposed
to the one that absorbs the $b$-quark spectator) is a small size
colour-singlet object, e.g. either a fast light meson
($\pi$, $\varrho$, $K$, $K^*$) or a $J/\Psi$. 
For the special case of the ratio
$\Gamma(B\to D^{*}\pi)/\Gamma(B\to D\pi)$ the perturbative
corrections to naive factorization have been evaluated in \cite{PW}
using a formalism similar to the one described above.
Note that factorization
cannot be justified in this way if the emitted particle is a
heavy-light meson ($D^{(*)}$), which is not a compact object
and has strong overlap with the remaining hadronic environment.
An important general caveat is the question 
of power corrections, which is relevant for assessing the accuracy
in phenomenological applications.
Moreover, (\ref{qfac}) has so far been demonstrated to one loop.
Clearly, more theoretical work is necessary to establish this
result at higher orders.
For more details on QCD factorization in $B$ decays see \cite{BBNS}
and the contribution of M. Beneke in these proceedings.

\subsection{$B\to J/\Psi K_S$ and $B\to\pi^+\pi^-$}

$B\to J/\Psi K_S$ is the prototype decay for the study of CP violation
in the $b$ sector. It is one of the rare cases where good experimental
feasibility and a clear signature are matched by exceedingly small
theoretical uncertainties.
For the latter property the decisive reason is the simple structure
of the amplitude, which has the form 
\begin{equation}\label{abpsik}
A(B\to\Psi K_S)=|A|e^{i\delta_s}e^{i\phi_w}
(1+{\cal O}(\lambda^2\cdot\mbox{penguin}))
\end{equation}
and where the small ${\cal O}(\lambda^2)$ contribution is negligible.
The amplitude is then governed by a single weak phase $\phi_w$.
As a consequence there is practically no direct CP violation.
Moreover, all hadronic uncertainties, related to the single factor
$|A|{\rm exp}(i\delta_s)$, cancel in the time-dependent mixing
induced CP asymmetry
\begin{eqnarray}\label{acppsiks}
{\cal A}_{CP} &=&
\frac{\Gamma(B(t)\to\Psi K_S)-\Gamma(\bar B(t)\to\Psi K_S)}{
      \Gamma(B(t)\to\Psi K_S)+\Gamma(\bar B(t)\to\Psi K_S)} 
 \nonumber \\
&=& -\sin 2\beta \, \sin\Delta M t
\end{eqnarray}
which allows for a simple and direct interpretation in terms of
the CKM angle $\beta$. Current constraints on the unitarity triangle
(see sect. 3) lead to the Standard Model expectation of 
$\sin 2\beta=0.71\pm 0.14$ \cite{AJB99}.
One of the most interesting recent developments in CP violation
has been the CDF measurement \cite{S2B}
\begin{equation}\label{s2bcdf}
\sin 2\beta=0.79^{+0.41}_{-0.44}\qquad \mbox{(CDF)}
\end{equation}
based on the full run-I data sample consisting of about
400 (untagged) $B\to J/\Psi K_S$ decays. The result has been made 
possible by a careful combination of several different tagging
strategies (see talk by G. Bauer, these proceedings). 
This $2$-$\sigma$ measurement provides
the first evidence for CP violation with $B$ mesons.
It excludes negative values of $\sin 2\beta$ at a confidence level
of $93\%$. Previous analyses were reported in \cite{S2Bx}.
The determination of $\sin 2\beta$ will be further pursued by several
experiments (BaBar, Belle, Hera-B, CDF (run II)) and exciting
new results can be expected in the near future.

A complementary source of information on CKM quantities is the
decay  $B\to\pi^+\pi^-$. As for $B\to J/\Psi K_S$, the final state
is an eigenstate of CP and an interesting CP asymmetry can be induced
by $B$--$\bar B$ mixing. However, now the amplitude has two
components
\begin{equation}\label{abpipi}
A(B\to\pi^+\pi^-)=V^*_{ub} V_{ud} T + V^*_{cb} V_{cd} P
\end{equation}
where both are of the same order in the Wolfenstein expansion and
have different weak phases 
($V^*_{ub} V_{ud}\sim\lambda^3\exp(i\gamma)$,
 $V^*_{cb} V_{cd}\sim\lambda^3$). For this reason the hadronic 
dynamics will not cancel completely when forming the CP asymmetry,
but will enter the observable in the ratio $P/T$, expected to
be about $10$ -- $20\%$ in modulus. The time dependent asymmetry
is then more complicated
\begin{equation}\label{acppipi}
{\cal A}_{CP}(B\to\pi^+\pi^-)=C\cos\Delta M t+ S\sin\Delta M t
\end{equation}
Only when $P/T\to 0$ do the coefficients reduce to a simple form,
$C=0$, $S=-\sin 2\alpha$.
In this case the asymmetry would be a direct measure of 
$\sin 2\alpha$, but the penguin amplitude $P$ cannot a priori be
neglected.
Various proposals exist to deal with this complication and
still make use of the information from $B\to\pi^+\pi^-$.

A well-known strategy is the isospin analysis of Gronau and London
\cite{GL}, a classical example for the use of flavour symmetry
to eliminate hadronic uncertainties.
In the case at hand this method requires the measurement of
$B^\pm\to\pi^\pm\pi^0$ and $B/\bar B\to\pi^0\pi^0$. Because of
the small expected branching ratio and the experimentally 
problematic final state of the latter decay, it appears to be
extremely difficult to apply this idea in practice.
Another option is to relate CP violation measurements in
$B\to\pi^+\pi^-$ with the corresponding observables in
$B_s\to K^+K^-$ and to use the fact that the hadronic dynamics of
these processes is related by U-spin symmetry
(the subgroup of flavour SU(3) acting on ($d$, $s$))
\cite{DSM,RFU1}. 
Finally, one may try to estimate
$P/T$ theoretically, which would allow us to extract
$\sin 2\alpha$ from a measurement of (\ref{acppipi}) alone 
(\cite{BBNS} and M. Beneke, these proceedings).

\subsection{$B\to\pi K$ and $\gamma$}

Decays of $B$ mesons into a pair of light pseudoscalars play a
central role in exploring CP violation. 
The recent CLEO discovery 
of such modes, which have small branching ratios $\sim 10^{-5}$,
has therefore attracted considerable interest.
In fact, experimental access to these processes opens new avenues
in the search for CP violating phenomena outside the kaon system,
for the determination of CKM parameters and for probing physics beyond 
the Standard Model.
Among the results reported by the CLEO collaboration
\cite{KWO}, the following are especially relevant for the present
discussion (for additional information see the talk by M. Artuso,
these proceedings)
\begin{equation}\label{bbpkc}
B(B\to\pi^\pm K^\mp)=(18.8^{+2.8}_{-2.6}\pm 1.3)\cdot 10^{-6}
\end{equation}
\begin{equation}\label{bbpckn}
B(B^\pm\to\pi^\pm K)=(18.2^{+4.6}_{-4.0}\pm 1.6)\cdot 10^{-6}
\end{equation}
\begin{equation}\label{bbpnkc}
B(B^\pm\to\pi^0 K^\pm)=({12.1^{+3.0+2.1}_{-2.8-1.4}})\cdot 10^{-6}
\end{equation}
\begin{equation}\label{bbpp}
B(B\to\pi^+\pi^-)=(4.7^{+1.8}_{-1.5}\pm 0.6)\cdot 10^{-6}
\end{equation}
\begin{equation}\label{bbpcpn}
B(B^\pm\to\pi^\pm\pi^0) < 12 \cdot 10^{-6}
\end{equation}
These measurements are still quoted as an average over charge
conjugated modes (e.g. $B^+\to\pi^0 K^+$ and $B^-\to\pi^0 K^-$).
Measured separately, a difference in branching ratio for these
conjugated processes would reveal CP violation.

To give an illustration of how the above results can, in principle,
be used to constrain CKM parameters, let us first consider the
ratio
\begin{equation}\label{rbpk}
R=\frac{B(B\to\pi^\pm K^\mp)}{B(B^\pm\to\pi^\pm K)}
\end{equation}
The following method was first suggested by Fleischer and Mannel
\cite{FM} and triggered a lively debate in the subsequent literature.
The decay mode $B^0\to\pi^- K^+$ proceeds at the quark level as
$\bar b(d)\to\bar su\bar u(d)$, which can be generated by tree-level
transitions (simple $W$-exchange $\bar b\to\bar su\bar u$ at zeroth
order in strong interactions) or by penguin contributions.
The amplitude has therefore two components, `tree' $t$ and `penguin' $p$,
as can also be read off from (\ref{heffb}):
\begin{equation}\label{abpkc}
A(B^0\to\pi^- K^+)=\lambda^2 e^{i\gamma} t+ p
\end{equation}
The weak phase of the tree contribution is
$\arg(V^*_{ub}V_{us})=\gamma$, the one of the penguin
$\arg(V^*_{ub}V_{us})\simeq 0$, in usual phase conventions.
By contrast the process $B^+\to\pi^+ K^0$ comes from
$\bar b(u)\to\bar sd\bar d(u)$ and has only a penguin contribution,
related to the previous one by isospin symmetry. Thus one expects
\begin{equation}\label{abpckn}
A(B^+\to\pi^+ K^0)\approx p
\end{equation}
neglecting electroweak penguins and rescattering effects for the moment.
Assuming (\ref{abpkc}) and (\ref{abpckn}), and the corresponding 
relations for the CP conjugated modes (where $\gamma\to -\gamma$),
it is straightforward to show that $\cos\gamma$ has to satisfy the
bound \cite{FM}
\begin{equation}\label{fmb}
|\cos\gamma|\geq\sqrt{1-R}
\end{equation}
independently of the hadronic quantity $p/t$. The
Fleischer-Mannel bound in its original form (\ref{fmb}) is useful
if $R<1$. In this case the bound excludes in the ($\varrho$, $\eta$)
plane an angular region centered around the positive $\eta$-axis.
While earlier results from CLEO seemed to indicate a low value of
$R$, the most recent numbers (\ref{bbpkc}), (\ref{bbpckn}) are
well compatible with $R\approx 1$. Nevertheless the present
simple example is useful to illustrate the basic idea behind
similar, more general approaches that could still be used to
extract information from $R$ \cite{BF,RF1}.
A further comment is in order concerning the bound (\ref{fmb}).
As mentioned above, the simple form of (\ref{abpckn}), leading to
(\ref{fmb}), neglects rescattering effects. At the quark level those
proceed through 
$(u)\bar b\to (u)\bar uu\bar s\to (u)\bar dd\bar s$,
where $u\bar u$ rescatters into $d\bar d$, similarly to the
penguin contraction shown in fig. \ref{fig:bme}. 
From the $\bar b\to\bar u$ transition a component with weak phase 
$\gamma$ is thus induced also in (\ref{abpckn}), modifying the above
argument.
In the partonic picture this effect is expected to be very small.
The validity of this description has subsequently been the subject
of a controversial discussion in the literature \cite{CONT}.
The recent developments in the theory of hadronic $B$-decays
\cite{BBNS} lend support to a quark-level description and the
original assumption of a small rescattering contribution to
$B^+\to\pi^+ K^0$. Correspondingly, the $B^+\to\pi^+ K^0$ amplitude
would be dominated by a single weak phase and direct CP violation
in the channels $B^\pm\to\pi^\pm K$ would be negligible. This
can be tested experimentally.

Another strategy to constrain $\gamma$ was discussed by
Neubert and Rosner \cite{NR,NEU}, 
based on earlier work in \cite{GRL,DHE}.
Here one starts from the ratio
\begin{equation}\label{rsbpk}
R^{-1}_*=\frac{2 B(B^\pm\to\pi^0 K^\pm)}{B(B^\pm\to\pi^\pm K)}
\end{equation}
Still neglecting rescattering effects, but incorporating dominant
electroweak penguin contributions using SU(3) flavour symmetry,
\cite{NR} obtain a bound of the form
($\delta_{EWP}\approx 0.7$)
\begin{equation}\label{nrb}
|\cos\gamma-\delta_{EWP}|>
\frac{\sqrt{B(B^\pm\to\pi^\pm K)}}{0.38\sqrt{B(B\to\pi^\pm\pi^0)}}
\left(1-\sqrt{R_*}\right)
\end{equation}
that can be translated into a bound on $\gamma$.

These ideas can be generalized to systematic analyses of
$B\to\pi K$ decays and the determination of $\gamma$. Theoretical
issues that can limit the
practical usefulness of a given approach, and that need to be addressed
in this context, are the role of electroweak penguins,
rescattering effects (final state interactions (FSI)) and corrections
to SU(3) flavour symmetry. Additional information can be gained by
combining $B\to\pi K$ with other modes, such as $B^+\to\pi^+\pi^0$,
$B\to K^+K^-$ and using results on direct CP violation, which should 
become available in the future. General analyses of this type
have been presented in \cite{BF,NEU,GPY,AD}.
Further theoretical insight into the dynamics of exclusive
nonleptonic decays in the heavy-quark limit, discussed above,
could be useful for a successful implementation of this program.

\subsection{Other Strategies}

To illustrate the variety of opportunities for probing CP
violation with $B$ decays we briefly present a selection of
further strategies that have been proposed in the literature.
We concentrate on approaches with good control of
theoretical uncertainties.

A clean extraction of $\gamma$ can in principle be obtained
from $B^\pm\to D_{CP}K^\pm$, $B^\pm\to D^0K^\pm$ and
$B^\pm\to \bar D^0K^\pm$ decays ($D_{CP}$ denotes a neutral
$D$ meson seen in a final state that is an eigenstate of CP).
The method was proposed in \cite{GW} and refined in
\cite{ADS}.

Another interesting approach makes use of $B_s\to D^\pm_s K^\mp$.
From a time-dependent analysis one can determine $\sin^2\gamma$
\cite{ADK}. As in the previous example, the underlying
quark-level transitions are $b\to u\bar cs$, $b\to c\bar us$.
Hence, no penguin contributions are possible.
Another advantage of this strategy are the sizable branching
ratios ($\sim 10^{-4}$), while a disadvantage is the need to resolve
the rapid time oscillations in the $B_s$--$\bar B_s$ system.

$B_s$ mesons can also be used to study the mixing induced
CP asymmetry ${\cal A}_{CP}(B_s\to\Psi\phi)$, which is a $B_s$
analog of ${\cal A}_{CP}(B_d\to\Psi K_S)$. Instead of $\sin 2\beta$,
${\cal A}_{CP}(B_s\to\Psi\phi)$ measures $2\lambda^2\eta\approx 3\%$.
The asymmetry is thus very small in the Standard Model.
It is therefore also interesting as a probe of New Physics
in $B_s$--$\bar B_s$ mixing, which could give much larger effects.

Additional ways of exploiting the rich information from
$B$ meson decays have been discussed by R. Fleischer at this
conference.
One of these possibilities is the combined use of 
$B_d\to\pi^+\pi^-$ and $B_s\to K^+ K^-$ \cite{DSM,RFU1}, 
already briefly mentioned above.
These processes are transformed into each other by simply exchanging
$d$ and $s$ quarks. The hadronic dynamics governing the two decays
is therefore related by U-spin symmetry, the subgroup of flavour
SU(3) operating on $d$ and $s$.
The time-dependent CP asymmetries of 
$B_d\to\pi^+\pi^-$ and $B_s\to K^+ K^-$ offer enough information to
determine, in the U-spin symmetry limit, the CKM angles $\beta$ and 
$\gamma$ independently of further hadronic input as emphasized in
\cite{RFU1}. There it is also shown how the impact of potential
U-spin breaking might be controled, for instance by using
$\beta$ (which will be well measured in the near future)
as an input to the analysis. The approach looks promising for the
LHC era, where $B_s\to K^+ K^-$ should be well accessible.

U-spin symmetry may also be exploited for probing $\gamma$
with combined information from 
$B_s\to J/\Psi K_S$ and $B_d\to J/\Psi K_S$ \cite{DSM,RFU2}.
Further options for extracting $\beta$ and $\gamma$ are provided by
investigating time-dependent angular distributions in
$B_d\to J/\Psi\varrho$ and $B_s\to J/\Psi\phi$ \cite{RF2}.

Recent discussions on New Physics effects in 
$B$ decays and CP violation are given in \cite{FMA,GNK,BFMA,BAFL}.

Besides the standard use of purely exclusive modes, the 
possibility of studying CP violation in semi-inclusive decays
has been proposed in the literature. In particular the
processes $B\to X_s\phi$ \cite{DEHT}, $B\to K^{(*)}X$ \cite{BDHP}
and $b\to\pi^- u$, $b\to K^- u$ \cite{AS} have been discussed.
It is also possible to search for CP violation in fully inclusive
$B$ decays, to which we turn in the following section.

\subsection{Inclusive CP Asymmetries}

CP violation in the $B_d$--$\bar B_d$ mixing matrix, measured by
$a\equiv{\rm Im}(\Gamma_{12}/M_{12})$, is one example of an inclusive
CP asymmetry. It arises from a relative phase between the mass-matrix 
element $M_{12}$ in $B$--$\bar B$ mixing and the off-diagonal
element in the $B$--$\bar B$ decay rate matrix
$\Gamma_{12}=\sum_f\langle B|f\rangle\langle f|\bar B\rangle$,
an inclusive quantity. $\Gamma_{12}$ can be computed using
the heavy-quark expansion (HQE) \cite{CCG,BIG}, 
exploiting the fact that
$m_b\gg\Lambda_{QCD}$ and assuming local quark-hadron duality.
The resulting Standard Model expectation is 
$a\stackrel{<}{_\sim} 10^{-3}$ for $B_d$ mesons (and smaller still for
$B_s$).
The effect could be enhanced by New Physics to typically
$a\sim 10^{-2}$ \cite{CW}.

The quantity $a$ can be measured using a lepton charge asymmetry
in time-dependent $B$ decay
\begin{equation}\label{alta}
{\cal A}_l=\frac{\Gamma(\bar B(t)\to l^+)-\Gamma(B(t)\to l^-)}{
                 \Gamma(\bar B(t)\to l^+)+\Gamma(B(t)\to l^-)}=a
\end{equation}
Using a similar method ALEPH finds (B. Petersen, these proceedings)
\begin{equation}\label{aaleph}
a=-0.05\pm 0.03\pm 0.01
\end{equation}
which is compatible with zero at a level of sensitivity of
a few percent.
The same quantity $a$ can also be determined from the tagged and
time-dependent totally inclusive $B$ decay asymmetry
\cite{BBD2}
\begin{eqnarray}\label{aall}
{\cal A}_{all}(t) &=& 
  \frac{\Gamma(B(t)\to\mbox{all})-\Gamma(\bar B(t)\to\mbox{all})}{
        \Gamma(B(t)\to\mbox{all})+\Gamma(\bar B(t)\to\mbox{all})}
\nonumber \\
&=& a \left[\frac{x}{2}\sin\Delta M t-\sin^2\frac{\Delta M t}{2}\right]
\end{eqnarray}
where $\Delta M$ is the mass difference between eigenstates in the
$B$--$\bar B$ system, $x=\Delta M/\Gamma$ (with $\Gamma$ the
total decay rate) and `all' denotes all possible final states.
Using this method $a$ has been measured to be
(\cite{ALEPHA,OPALA,SLDA,DELPHIA} and 
B. Petersen, R. Hawkings, these proceedings)
\begin{eqnarray}\label{aallex}
a &=& -0.006\pm 0.043^{+0.011}_{-0.009}\qquad\mbox{(ALEPH)}
\nonumber\\
a &=& 0.005\pm 0.055 \pm 0.013\qquad\mbox{(OPAL)}
\nonumber\\
a &=& -0.04\pm 0.12 \pm 0.05\qquad\mbox{(SLD)}
\nonumber\\
a &=& -0.022\pm 0.030 \pm 0.011\qquad\mbox{(DELPHI)}
\nonumber
\end{eqnarray}
in agreement with (\ref{aaleph}). Also direct CP violation may be
looked for inclusively. The most up-to-date analysis of such
asymmetries in $B$ decays to charmless final states $X$
\begin{equation}\label{adirinc}
{\cal A}_{dir,incl}=\frac{\Gamma(B^+\to X)-\Gamma(B^-\to X)}{
                          \Gamma(B^+\to X)+\Gamma(B^-\to X)}
\end{equation}
is given in \cite{LNO}. There it is estimated that
\begin{eqnarray}\label{adinc12}
{\cal A}_{dir,incl}(\Delta S=0) &=&\left(2.0^{+1.2}_{-1.0}\right)\%
\\
{\cal A}_{dir,incl}(\Delta S=1) &=&\left(-1.0\pm 0.5\right)\%
\end{eqnarray}
for transitions with $\Delta S=0$ and $\Delta S=1$, respectively.

Finally, flavour-specific inclusive decays (such as
$b\to u\bar ud$) may also be used to construct mixing-induced
CP asymmetries \cite{BBD2}.

\section{Summary}

At present it is still true that CP violation has been
firmly established only in a few decay channels of the long-lived
neutral kaon, namely
$K_L\to\pi\pi$, $K_L\to\pi l\nu$, $K_L\to\pi^+\pi^-\gamma$ and
$K_L\to\pi^+\pi^- e^+ e^-$. The main effect is described by a
single complex parameter $\varepsilon$, which is well measured
and whose size is in good agreement with the Standard Model.
Despite some hadronic uncertainties, this agreement is non-trivial
and naturally accomodates what is known from independent sources
for $m_t$, $V_{cb}$ or $V_{ub}$. Simultaneously, what is thus
learned from $\varepsilon$ in the kaon system implies that large
mixing-induced CP violation should be seen in the $B$ sector,
$\sin 2\beta\approx 0.7\pm 0.1$.

Although continuous progress has occurred over the years, leading
to important refinements, the basic empirical knowledge on
CP nonconservation has nevertheless remained rather limited.
This situation is expected to change considerably in the near future.
It is probably no exaggeration to say that we are at the beginning
of a new era in flavour physics and CP violation. In fact this
new phase is already marked by several highlights that were
recently reported and presented at this conference:

\begin{itemize}
\item 
Direct CP violation is now, for the first time, firmly established
in $K\to\pi\pi$ decays (KTeV, NA48).
\item
T violation has been demonstrated by the CPLEAR 
collaboration in $K\leftrightarrow\bar K$
transitions.
\item
The rare decay $K_L\to\pi^+\pi^- e^+e^-$ has been observed and 
found to exhibit a large, T-odd, CP violating asymmetry in the decay
angular distribution (KTeV, NA48), in nice agreement with theoretical
predictions.
\item
Brookhaven experiment E871 has measured 
$B(K_L\to e^+e^-)=(8.7^{+5.7}_{-4.1})\cdot 10^{-12}$, the smallest
decay branching fraction yet observed.
\item
Brookhaven experiment E787 has seen one event of the `golden'
decay $K^+\to\pi^+\nu\bar\nu$, corresponding to a branching
ratio of $(1.5^{+3.5}_{-1.3})\cdot 10^{-10}$. The continuing search is
approaching Standard Model sensitivity, a development that opens the
way to high precision flavour physics.
\item
First evidence for large CP violation in the $B$ sector is reported
from CDF, providing a $2\sigma$ signal for the asymmetry in
$B/\bar B\to\Psi K_S$ and $\sin 2\beta=0.79^{+0.41}_{-0.44}$.
\item
The CLEO experiment at Cornell has performed the first measurements
of the branching fractions of several rare ($\sim 10^{-5}$),
nonleptonic $B$ decays, among them
$B\to K^+\pi^-$, $K^+\pi^0$, $K^0\pi^+$, $K^0\pi^0$ and
$B\to\pi^+\pi^-$.
This class of processes will undoubtedly play an important role
for the phenomenology of CP violation in the coming years.
\end{itemize}

These highlights give us a first glimpse of the rich prospects
offered by the world-wide program of precision flavour physics.
It is obvious that for a comprehensive view both dedicated rare
kaon experiments as well as detailed studies of $B$ decays
are necessary. Together they will provide the complementary information
that is crucial for a decisive test of the CKM paradigm and for
probing what may lie beyond. Other sources of information, as
$D$ meson decays, electric dipole moments, etc., could well harbour
some surprizes and should also be pursued.

The $B$ factory experiments BaBar at SLAC and Belle at KEK, and the
KLOE experiment at the Frascati $\phi$ factory have already
started. In the near future they will be joined by HERA-B, CLEO-III,
and the upgraded CDF and D0 detectors at Fermilab for 
Tevatron Run-II, while several kaon experiments are still ongoing
at CERN (NA48), Fermilab (KTeV), Brookhaven and KEK.
For the future, second generation $B$ precision experiments
(LHC-b at CERN, BTeV at Fermilab) are taking shape, supplemented
by ambitious rare $K$ decay projects
at Brookhaven (E926 and E949), Fermilab (KAMI and CKM) and
KEK (JHF).
Theoretical efforts are continuing to accompany the experimental
and technological progress. The weak decay Hamiltonians are now
routinely treated at the next-to-leading order in QCD, theoretically
clean observables have been identified, phenomenological strategies
are being proposed, various field theoretical methods 
(lattice QCD, QCD sum rules, $1/N_c$ expansion, chiral perturbation theory,
factorization) improve our understanding of weak hadronic processes.

The years to come therefore clearly provide exciting opportunities
for the phenomenology of weak decays and CP violation, resulting from
a close interplay of experimental discovery
and theoretical interpretation.

\section*{Acknowledgements}

It is a pleasure to acknowledge many
illuminating discussions with M. Beneke and
G. Isidori. I am also grateful to A. J. Buras,
G. D'Ambrosio, J. Donoghue, I. Dunietz, R. Fleischer, T. Hurth,
Z. Ligeti, G. Martinelli, U. Nierste, M. Peskin and A. Soni
for interesting conversations.
Thanks are due to T. Hurth for a critical reading of the manuscript.

\end{document}